\documentclass[a4paper]{article}
\usepackage[margin=1.0in]{geometry}
\usepackage{amsmath}
\usepackage{amsthm}
\usepackage{amsfonts}
\usepackage{amssymb}
\usepackage{jheppub}
\usepackage[table]{xcolor}
\usepackage{colortbl}
\usepackage{tikz}
\usepackage{ulem}

\usepackage{multirow}
\usepackage{siunitx}

\usetikzlibrary{decorations.pathmorphing}
\usetikzlibrary{shapes}
\usetikzlibrary{shapes.geometric}
\usetikzlibrary{decorations.pathreplacing}
\usepackage[force]{feynmp-auto}
\usepackage{tabularx}
\usepackage{pbox}
\usepackage{slashed}
\usepackage{hyperref}
\hypersetup{
    linktocpage=true,
    colorlinks=true,
    linkcolor=blue,
    citecolor=red,
    pdftitle={Title},
    bookmarks=true,
    pdfpagemode=FullScreen,
}

\newcommand{\mpl}{M_{\text{pl}}}

\newcommand{\gsim}{\lower.7ex\hbox{$\;\stackrel{\textstyle>}{\sim}\;$}}
\newcommand{\lsim}{\lower.7ex\hbox{$\;\stackrel{\textstyle<}{\sim}\;$}}

\usepackage{xcolor}
\usepackage{soul}

\begin{document}
\title{\begin{center}
		\LARGE{\textbf{Cosmological Particle Production \\ and Pairwise Hotspots on the CMB}}
	\end{center}
}
\author[a,b,c]{Jeong Han Kim,}
\author[d,e]{Soubhik Kumar,}
\author[f]{Adam Martin,}
\author[f]{Yuhsin Tsai}

\affiliation[a]{Department of Physics, Chungbuk National University, Cheongju, Chungbuk 28644, Korea}
\affiliation[b]{Center for Theoretical Physics of the Universe,
Institute for Basic Science, Daejeon 34126, Korea}
\affiliation[c]{Korea Institute for Advanced Study (KIAS), School of Physics, Seoul 02455, Korea}
\affiliation[d]{Berkeley Center for Theoretical Physics, Department of Physics,
University of California, Berkeley, CA 94720, USA}
\affiliation[e]{Theoretical Physics Group, Lawrence Berkeley National Laboratory, Berkeley, CA 94720, USA}
\affiliation[f]{Department of Physics, University of Notre Dame, IN 46556, USA}
\emailAdd{jeonghan.kim@cbu.ac.kr}
\emailAdd{soubhik@berkeley.edu}
\emailAdd{amarti41@nd.edu}
\emailAdd{ytsai3@nd.edu}
\abstract{
Heavy particles with masses much bigger than the inflationary Hubble scale $H_*$, can get non-adiabatically pair produced during inflation through their couplings to the inflaton. If such couplings give rise to time-dependent masses for the heavy particles, then following their production, the heavy particles modify the curvature perturbation around their locations in a time-dependent and scale non-invariant manner. This results into a non-trivial spatial profile of the curvature perturbation that is preserved on superhorizon scales and eventually generates localized hot or cold spots on the CMB. We explore this phenomenon by studying the inflationary production of heavy scalars and derive the final temperature profile of the spots on the CMB by taking into account the subhorizon evolution, focusing in particular on the parameter space where pairwise \textit{hot} spots (PHS) arise. When the heavy scalar has an $\mathcal{O}(1)$ coupling to the inflaton, we show that for an idealized situation where the dominant background to the PHS signal comes from the standard CMB fluctuations themselves, a simple position space search based on applying a temperature cut, can be sensitive to heavy particle masses $M_0/H_*\sim\mathcal{O}(100)$. The corresponding PHS signal also modifies the CMB power spectra and bispectra, although the corrections are below (outside) the sensitivity of current measurements (searches).
}
\maketitle

\section{Introduction}
Cosmic inflation (for a review see~\cite{Baumann:2009ds}) is one of the leading paradigms to explain the generation of primordial fluctuations that seed the cosmic microwave background (CMB) anisotropies and the large-scale structure (LSS) of the Universe. Similar to particle collider experiments, the inflationary era, characterized by the Hubble expansion parameter $H_*$, generically provides an environment to produce new particles with masses $\sim H_*$ which can in turn imprint their signatures on primordial fluctuations. Given $H_*$ during inflation can as high as $5\times 10^{13}$~GeV~\cite{Planck:2018jri}, this provides us with a unique chance to probe very high-energy physics operating at scales $\gtrsim H_*$. 

During inflation, the expanding spacetime leads to a time-dependent Hamiltonian, and particles with masses $m \sim H_*$ can be produced non-adiabatically even when starting with the Bunch-Davies vacuum at very early times~\cite{Birrell:1982ix}. While such minimal non-adiabatic production amplitude gets exponentially suppressed as $e^{-\pi m/H_*}$ for $m\gg H_*$, the slowly rolling inflaton field itself can assist in producing even heavier particles if a direct coupling exists between the inflaton and the heavy particles. The produced particles can (a)~generate a non-trivial spatial correlation between curvature perturbations~$\langle\zeta(\vec{x}_1) \zeta(\vec{x}_2) \cdots\rangle$ at different locations or, (b)~if sufficiently heavy, they can directly modify the curvature perturbation~$\langle\zeta(\vec{x})\rangle$ around their individual locations, as will be explained below. In the absence of isocurvature fluctuations, curvature perturbations $\zeta$ then remain conserved on superhorizon scales by causality~\cite{Weinberg:2003sw, Wands:2000dp}. As they re-enter the horizon, the sub-horizon physics, such as the baryon acoustic oscillations (BAO) or the structure formation process, further process these signals and the imprints of the heavy particles are recorded on the CMB and the LSS.

The phenomena of inflationary particle production, along with the associated signatures in the power spectrum and non-Gaussianity of $\zeta$ (option~(a) mentioned above), has been studied extensively in the literature. A relatively recent example of this is the ``Cosmological Collider Physics'' program~\cite{Chen:2009zp, Arkani-Hamed:2015bza} that aims to detect \textit{on-shell} mass-spin signatures of heavy particles via non-analytic momentum dependence and angular dependence of cosmological correlators. In particular, it has been shown recently~\cite{Chen:2018xck, Hook:2019zxa, Wang:2020ioa, Bodas:2020yho, Sou:2021juh} that particle masses $m\lesssim \sqrt{\dot{\phi}_0}$, where $\sqrt{\dot{\phi}_0}\approx 60 H_*$~\cite{Planck:2018jri} is the slowly rolling inflaton velocity, can give rise to distinctive oscillatory signatures in primordial non-Gaussianity. For high-scale inflation, such a scale $\sim 60 H_*$ is interesting from a particle physics point of view since it can be close to the scale of Grand Unification. 

Motivated by this, in this work we ask what kind of mechanisms can produce even heavier particles with their typical masses~$m\gg\sqrt{\dot{\phi}_0}$, and if they can lead to novel phenomenology. A shift-symmetric, derivative coupling of the inflaton to the heavy particles involves factors of $\partial_\mu\phi$ and therefore, $\sqrt{\dot{\phi}_0}$ is the maximum energy budget that can be used to produce heavy particles and correspondingly, production for masses~$m\gg\sqrt{\dot{\phi}_0}$ is extremely suppressed. However, if we allow shift-symmetry breaking interactions, this conclusion is avoided. A simple example of this involves a heavy particle whose mass is dependent on the inflaton field value. Then as the inflaton rolls along its potential, the heavy field mass may pass through a minimum, and depending on that efficient particle production can occur at that instant while it shuts off at other times.

The above mechanism with shift-symmetry violation has been noted before, see e.g.~\cite{Chung:1998zb, Chung:1999ve, Kofman:2004yc, Romano:2008rr, Barnaby:2009mc} and~\cite{Chluba:2015bqa} for a review. Here we incorporate two differences. First, the minimum of the heavy field mass is always $\gg H_*$, similar to~\cite{Flauger:2016idt} and therefore, the slow-roll inflaton trajectory need not be significantly perturbed. The number of produced heavy particles is also significantly smaller compared to scenarios where heavy particle masses pass through values $\lesssim H_*$. Second and more importantly, following their production, the heavy particles actually get significantly massive (and non-relativistic) such that they can directly modify the curvature perturbation by giving rise to a non-zero one-point function $\langle\zeta(\vec{x})\rangle$ around their almost static locations. These localized perturbations, after reentering the horizon, give rise to spots/structures localized in position space. Given such a localized nature of the signal that is not repeated across various scales in the sky, it is natural to wonder how a direct position space search for such signatures compares to the traditional approach involving power spectrum, bispectrum or trispectrum of density perturbations.

To investigate this, in this work we present a concrete scenario in which such signatures show up as pairwise hotspots (PHS) on the CMB map. We focus on an example involving a heavy scalar field with an inflaton-dependent mass. We compute the number density of the heavy scalars following their production, and this determines the number of hotspots that appear on the CMB surface after an appropriate volume dilution. For each individual heavy scalar, we also compute the associated curvature perturbation $\langle\zeta(\vec{x})\rangle$ during inflation and then propagate it forward through subhorizon evolution upon its horizon reentry. In particular, by taking into account the necessary Sachs-Wolfe and integrated Sachs-Wolfe effects, we obtain the final localized temperature profile corresponding to each heavy scalar. Owing to their non-adiabatic production from time-dependent vacuum, the heavy particles come in pairs and therefore we finally get pairwise signals on the CMB sky. As we will show later, the two hotspots of a given pair generally overlap with each other.

To make the comparison between position and momentum space searches mentioned above, we study both how the PHS can give rise to localized distortions in the CMB TT-spectrum and also localized signals on the position space. For the position space study, we simulate the effects of PHS using the {\tt HEALPix}~\footnote{\url{https://healpix.sourceforge.io/}} package and perform a simplified ``cut-and-count'' analysis to identify the PHS from the background of standard CMB fluctuations. For the power spectrum based search, we numerically compute the CMB TT-spectrum in the presence of PHS using \texttt{CLASS}~\cite{Lesgourgues:2011re,Blas:2011rf}. Our simplified analysis indeed demonstrates situations where a position space cut-and-count analysis performs better than the power spectrum based search. In particular, we show that the position space search may probe the minimum value of the scalar mass up to $\mathcal{O}(100)H_*$ -- comparable to the kinetic energy of the inflaton. At the same time, the search crucially relies on typical values of the heavy mass away from the minimum and that can be as big as $\mathcal{O}(10^4)H_*$. In this sense, this search is sensitive to very high mass scales and can go beyond Grand Unification scales as well. While we do not consider the presence of instrumental noise and astrophysical foregrounds, we also do not use the detailed morphology of the signal in our search that is expected to mitigate the former two contributions. We will address this important aspect in a future work. We also do not consider in detail the resulting non-Gaussianity signatures for which a dedicated template study would be needed since the signatures show up on narrow windows in momentum space. 

Some earlier literature have discussions on the production of heavy particles that relate to our study. The example model we use containing an inflaton-dependent scalar mass has similar behaviors to the examples studied in~\cite{Kofman:1997yn, Flauger:2016idt}. However, Ref.~\cite{Flauger:2016idt} and subsequently Ref.~\cite{Munchmeyer:2019wlh} focus on $N$-point correlation functions to study the signatures of periodic particle productions. In contrast, we focus on the \textit{position} space signatures of heavy particles produced at a \textit{single} instant of time during inflation. Therefore, in our set up the hotspots are characterized by a single scale and distortions in the momentum space $\zeta$-correlators show up on a specific scale without any periodicity. 
Regarding the calculation of the temperature profile of the hotspot, Ref.~\cite{Fialkov:2009xm} has a similar discussion although they assume the existence of pre-inflationary heavy particles that appear in our visible horizon. However, since inflation dilutes any pre-existing relic, one drawback of such an assumption is that the final number density of the hotspots strongly depends on the unknown amount of inflation we have had, strongly affecting the phenomenology. Our set up is free from this issue since particles are produced during inflation itself when the CMB-observable modes exit the horizon. Furthermore, the hotspots/defects they consider appear on much larger scales with different phenomenology.
The idea of generating the PHS signals on the CMB map was discussed in Ref.~\cite{Maldacena:2015bha} with the motivation of developing a quantum-entangled system for Bell's inequality test of the Universe's primordial fluctuations. Here in the context of a full model we also compute the number of such hotspots and study their properties. We also go beyond the computation of the primordial curvature perturbation and include the important effects of subhorizon transfer functions that are essential for the position space search.

The plan of the paper is as follows. In Sec.~\ref{sec.partprod} we describe our model and review some necessary aspects of particle production. We then compute how many hotspots can be produced while ensuring small backreaction to the slow-roll inflaton dynamics. In Sec.~\ref{sec.hspheno} we discuss the detailed phenomenology of such hotspots by taking into account the subhorizon evolution. In particular, we derive both the central temperature of the hotspots as a function of the comoving horizon size at the time of particle production, and the spatial profiles of individual hotspots. These profiles are then injected into mock CMB maps generated via \texttt{HealPix} as described in Sec.~\ref{sec.detstrat}. There we also outline our position space search strategy. After studying the effects of PHS on the CMB-TT power spectrum, we present our bounds on the mass-coupling parameter space. We conclude in Sec.~\ref{sec.concl}. In Appendix~\ref{app:profile}, we give some details of the ``in-in'' computation of the curvature perturbation due to a massive particle, while Appendix~\ref{app.img} contains some benchmark CMB images relevant for our study.

\section{Heavy particle production during inflation}\label{sec.partprod}
\subsection{Set-up}
We denote the background inflationary spacetime metric by,
\begin{align}
    ds^2 = -dt^2+a(t)^2d\vec{x}^2,
\end{align}
where $a(t)=e^{H_*t}$ is the scale factor given in terms of the inflationary Hubble scale $H_*$ which approximately remains constant during inflation. We will consider a scenario in which an otherwise ultra-heavy particle $\sigma$ momentarily becomes lighter as the inflaton passes through some field value, and is significantly produced only around that time. To model such events of particle production where the mass of the heavy field passes through a minimum, we consider the following Lagrangian,
\begin{align}\label{eq.heavylag}
\mathcal{L}\supset -\frac{1}{2}(\partial_\mu\sigma)^2 -\frac{1}{2} \left((g\phi-\mu)^2+M_0^2\right)\sigma^2.
\end{align}
This defines an inflaton-dependent effective mass for $\sigma$,
\begin{align}\label{eq.effmass}
M^2(\phi) = (g\phi-\mu)^2+M_0^2, 
\end{align}
that passes through its minimum at $\phi=\mu/g$. While in the context of more UV complete scenarios such as~\cite{Silverstein:2008sg, Flauger:2016idt}, and in scenarios with random mass functions~\cite{Green:2014xqa, Amin:2015ftc}, the full effective mass can have a more complicated dependence on $\phi$, around the time of particle production, we can still approximate $M^2(\phi)$ as a quadratic function of $\phi$ as in Eq.~\eqref{eq.effmass} quite generally. Using this parameterization, we will capture the physics of particle production through the parameters $g$ and $M_0$. One of the goals of this work is to use simulations of the CMB map to derive constraints on these two parameters. As we will see shortly, for the kind of signatures that we are looking for, we will typically need $1\lesssim g\lesssim 10$. While such values of $g$ can potentially give rise to large radiative corrections to the inflationary potential, in scenarios with additional symmetries, such as supersymmetry, the radiative corrections can be in control, see e.g.~\cite{Flauger:2016idt}.

\begin{figure}
    \centering
    \includegraphics[width = 5cm]{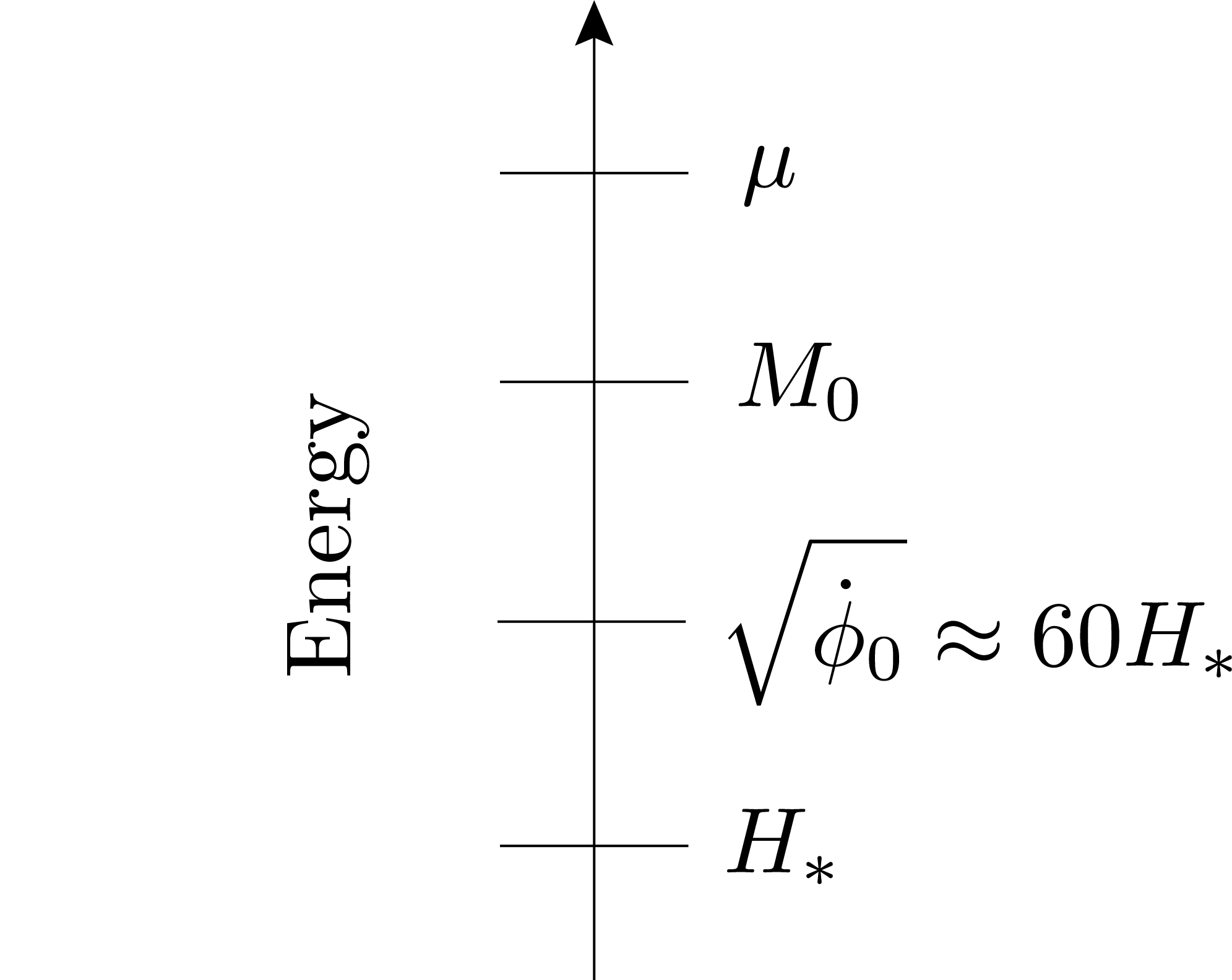}
    \caption{A summary of relevant mass scales considered in this work. At a typical point on the inflationary trajectory, the heavy field mass is $\sim\mu\gg \sqrt{\dot{\phi}_0}$, the slowly rolling inflaton velocity. However, the mass passes through a minimum $M_0\ll \mu$ during a single instant of (conformal) time $\eta_*$ and non-negligible particle production takes place for $M_0\gtrsim \sqrt{\dot{\phi}_0}$.}
    \label{fig:scales}
\end{figure}

Before getting into a detailed calculation, we briefly describe the qualitative feature of the above Lagrangian~\eqref{eq.heavylag} whose relevant scales are shown in Fig.~\ref{fig:scales}.
We consider scenarios in which at a generic point on the inflationary trajectory, $g\phi\sim \mu \gg M_0\gg H_*$ and hence, the $\sigma$-mass is very large and cosmological production $\sigma$ particles is exponentially suppressed and negligible. However, as the inflaton passes through the field value $\phi_* = \mu/g$, $M(\phi)$ becomes significantly smaller momentarily, $M(\phi)\approx M_0$.  As a result, for $M_0 \sim \sqrt{\dot{\phi}_0}$, where $\sqrt{\dot{\phi}_0}\approx 60 H_*$ is the slow-roll inflaton velocity determined by the scalar power spectrum~\cite{Planck:2018jri}, the kinetic energy of $\phi$ can be harnessed to produce $\sigma$-particles when $\phi\approx\phi_*$, corresponding to conformal time $\eta_*$. Here and in the following, we use `$*$' to denote quantities evaluated at the time particle production.
By construction, this mechanism of heavy particle production lasts for a short period of time and the imprints of the heavy particles would be most prominent on some particular $k$-modes that exit the horizon around $\eta_*$. Therefore, the associated signature would show up as a violation of scale invariance. Our aim is to devise detection strategies to isolate such violations. While these effects would be observable in both the CMB and the LSS, in this work, we mostly focus on the effects on the CMB, leaving the effects on LSS for a future study.

\subsection{Review of particle production}
For the sake of completeness, we now review how to compute particle production given Eq.~\eqref{eq.heavylag}. Our approach will be somewhat pedagogical and readers can skip to the final result in Eq.~\eqref{eq.beta2} if needed. We start with the equation of motion (EOM) for $\sigma$,
\begin{equation}
\sigma''-\frac{2}{\eta}\sigma'+\left(k^2+\frac{M^2(\eta)}{H_*^2\eta^2}\right)\sigma=0\,,
\end{equation}
written in terms of the conformal time $\eta\equiv \int dt/a(t)$. In the above and the following, $'$s denote derivatives with respect to $\eta$.
To further simplify the EOM into a harmonic oscillator form, we define
$u=\sigma/\eta$, in terms of which we get,
\begin{equation}\label{eq.eom}
u''+\left(k^2+\frac{M^2(\eta)/H_*^2-2}{\eta^2}\right)u\equiv u''+\omega^2(\eta) u=0\,,
\end{equation}
with a time-dependent frequency,
\begin{align}\label{eq.omega}
    \omega^2 = k^2+\frac{M^2(\eta)-2H^2}{H_*^2\eta^2}.
\end{align}
To estimate how large the effects of particle production and violation of adiabaticity is, we can compute $\omega'(\eta)/\omega^2$ 
\begin{align}
\frac{\omega'}{\omega^2}=\frac{(MM'\eta-M^2)H_*}{(k^2\eta^2H^2+M^2)^{3/2}}.
\end{align}
up to terms $\mathcal O(H_*^2/M^2_0)$ which we neglect. Let us now focus on our model where the time dependent mass is given in Eq.~\eqref{eq.effmass}. Denoting $\mu\equiv g\phi_*$ and using the fact that during inflation the homogeneous solution under the slow-roll approximation goes as,
\begin{equation}
    \phi-\phi_* = \dot{\phi}(t-t_*),
\end{equation}
we can rewrite the mass term as ($\eta\propto -e^{-Ht}$),
\begin{align}\label{timedepmass}
    M^2 = \frac{g^2\dot{\phi}^2}{H_*^2}\ln(\eta/\eta_*)^2 + M_0^2.
\end{align}
This gives a measure of adiabaticity violation,
\begin{align}
\Bigg|\frac{(MM'\eta-M^2)H_*}{M^3}\Bigg| = \Bigg|\frac{\left(\frac{g^2\dot{\phi}^2}{H_*^2}\ln(\eta/\eta_*)-\frac{g^2\dot{\phi}^2}{H_*^2}\ln(\eta/\eta_*)^2 - M_0^2\right)H_*}{(\frac{g^2\dot{\phi}^2}{H_*^2}\ln(\eta/\eta_*)^2 + M_0^2)^{3/2}}\Bigg|\simeq\begin{cases} 
   \frac{H_*}{M_0} & \text{if } \eta \approx \eta_* \\
   \frac{H_*^2}{g\dot{\phi}|\ln{(\eta/\eta_*)|}}       & \text{generic } \eta
  \end{cases},    
\end{align}
where we have assumed $M_0\ll g\dot{\phi}|\ln{(\eta/\eta_*)|}/H_*$ to get the last relation. We have also set $k=0$ to estimate the maximum amount of adiabaticity violation. This then shows that particle production is most efficient just around $\eta_*$ where the mass is minimized. The same conclusion can be obtained via the steepest descent method along the lines of~\cite{Chung:1998bt}.

Given this we will study the simplified equation of motion around $\eta_*$ to study particle production. Therefore, we approximate
\begin{align}
(g\phi-\mu)^2 \equiv g^2(\phi-\phi_*)^2\approx g^2\phi^{\prime 2}(\eta-\eta_*)^2.   
\end{align}
Using this Eq. \eqref{eq.eom} can be written around $\eta=\eta_*$ as,
\begin{align}\label{eq.simpleeom}
\frac{d^2u}{d\tau^2}+(\kappa^2+\tau^2)u=0,    
\end{align}
where
\begin{align}
\gamma^4 = \frac{g^2\phi^{\prime 2}}{H_*^2\eta_*^2}; \hspace{2em} \tau = \gamma\eta-\gamma\eta_*; \hspace{2em} \kappa^2 = \frac{k^2}{\gamma^2}+\frac{M_0^2-2H^2}{\gamma^2\eta_*^2H^2}.    
\end{align}
To estimate particle production, we will solve Eq.~\eqref{eq.simpleeom}. We follow the standard procedure (see e.g.,~\cite{Birrell:1982ix}) to estimate the Bogoliubov-$\beta$ coefficient by studying the amplitude of a negative frequency mode that develops after starting with a positive frequency solution at very early times. To this end, let us first obtain the adiabatic solutions for small and large times. For $\tau\rightarrow\infty$ the $\pm-$frequency solutions are,
\begin{align}
e^{\mp i\int d\tau \sqrt{\kappa^2+\tau^2}}\rightarrow e^{\mp \frac{i}{2}\tau^2}.
\end{align}
For $\tau\rightarrow -\infty$ on the other hand, $\pm-$frequency solutions read as,
\begin{align}
e^{\mp i\int d\tau \sqrt{\kappa^2+\tau^2}}\rightarrow e^{\pm  \frac{i}{2}\tau^2}.
\end{align}
The exact solutions to Eq. \eqref{eq.simpleeom} can be written as parabolic cylinder functions $W\left(-\frac{\kappa^2}{2},\pm\sqrt{2}\tau\right)$~\cite{NIST:DLMF,Kofman:1997yn}. 
To obtain $\beta$ we will look at the asymptotic behavior of $W$. It can be checked that the following linear combination gives the positive frequency solution at $\tau\rightarrow -\infty$,
\begin{align}
u(\kappa,\tau)=i\sqrt{\sigma} W\left(-\frac{\kappa^2}{2},+\sqrt{2}\tau\right)+\frac{1}{\sqrt{\sigma}}W\left(-\frac{\kappa^2}{2},-\sqrt{2}\tau\right),
\end{align}
where $\sigma=\sqrt{1+e^{-\pi \kappa^2}}-e^{-\pi\kappa^2/2}$. The Bogoliubov-$\beta$ coefficient can be read off from the negative frequency part of this solution as $\tau\rightarrow+\infty$ which is given by,
\begin{align}
u(\kappa,\tau\rightarrow+\infty) = \frac{2^{1/4}}{\sqrt{\tau}}\left(\left(\frac{i\sigma}{2}-\frac{i}{2\sigma}\right)e^{+\frac{i}{2}\tau^2}+\left(\frac{i\sigma}{2}+\frac{i}{2\sigma}\right)e^{-\frac{i}{2}\tau^2}\right).
\end{align}
Thus we have the standard Bogoliubov $\alpha$ and $\beta$ coefficients as,
\begin{align}
\alpha =\left(\frac{i\sigma}{2}+\frac{i}{2\sigma}\right),~~
\beta = \left(\frac{i\sigma}{2}-\frac{i}{2\sigma}\right).
\end{align}
It can be checked that they satisfy, $|\alpha|^2-|\beta|^2=1$ as required. Thus for a given $k$-mode we get~\cite{Kofman:1997yn,Flauger:2016idt},
\begin{align}\label{eq.beta2}
|\beta|^2 = e^{-\pi\kappa^2} = \exp\left({-\frac{\pi (k^2\eta_*^2H^2+M_0^2-2H^2)}{|g\dot{\phi}|}}\right) \sim e^{-\frac{\omega^2}{\dot{\omega}}}\bigg\rvert_{\rm max},
\end{align}
where we have approximated, $\phi^\prime \approx -\frac{\dot{\phi}}{H_*\eta_*}$. This implies that for any comoving momentum $k$, particle production is Boltzmann suppressed if the minimum value of the time-dependent mass of the particle, here $M_0$, is bigger than $\sqrt{g\dot{\phi}}$. This can be understood as coming from the adiabaticity violation factor $e^{-\frac{\omega^2}{\dot{\omega}}}$. Focusing on $k=0$ for simplicity, $\omega^2=(g\phi-\mu)^2+M_0^2$ is the time-dependent frequency. The quantity $\dot{\omega}/\omega^2$ parameterizes the violation of adiabaticity and $\dot{\omega}/\omega^2$ is maximized when $g\phi-\mu = M_0/\sqrt{2}\simeq M_0$. This implies at the time of particle production, $\omega\simeq M_0$ and $\dot{\omega}\simeq g\dot{\phi}$ which gives, $|\beta|^2\sim e^{-\frac{\omega^2}{\dot{\omega}}}\simeq e^{-M_0^2/(g\dot{\phi})}$. We also mention that Eq.~\eqref{eq.beta2} is only valid for $|\beta|^2<1$ and cannot be applied to scenarios where $M_0^2<2H^2$, which lie outside our parameter space of interest.

\subsection{Estimating particle production}
\label{sec:partprod}
We can now calculate the physical number density of produced particles at $\eta_*$ after a phase space integral over physical momenta $k_{\rm phys}=k|H_*\eta_*|$,
\begin{align}
n &= \frac{1}{2\pi^2} \int d k_{\text{phys}}k_{\text{phys}}^2 e^{-\frac{\pi k_{\text{phys}}^2}{|g\dot{\phi}|}} e^{-\frac{\pi(M_0^2-2H^2)}{|g\dot{\phi}|}}\nonumber\\ \label{eq.nodensity}
&=\frac{1}{8\pi^3}\left(g\dot{\phi}\right)^{3/2} e^{-\frac{\pi(M_0^2-2H^2)}{|g\dot{\phi}|}}.
\end{align}

As we will show in the following, each of the particles leads to a localized ``disturbance'' in the spacetime metric that can manifest itself as either a hot or a cold spot on the CMB. The hot/cold nature depends on the size of the disturbance and is primarily controlled by $\eta_*$. As we will detail in Sec.~\ref{sec.hspheno}, for the benchmark choices of $\eta_*$ we will focus on, the disturbances will be \textit{hot}spots on the CMB and therefore we will use this terminology from now on. 
\paragraph{Notation.} Except for places where $|\eta_*|$ appears, we will now absorb the minus sign in $\eta_*$ and use it to denote the size of the comoving horizon at the time of particle production, instead of the (negative) conformal time.

The number of such hotspots within the CMB surface, about $\eta_{0}=13.8$~Gyr~\cite{Aghanim:2018eyx} (conformal age of the Universe) distance away with a thickness $\Delta \eta_{\rm rec}$,
is given after an appropriate volume dilution factor,
\begin{align}\label{eq.nhotspots}
N_{\rm HS} &= \frac{1}{8\pi^3}\left(\frac{g\dot{\phi}}{H_*^2}\right)^{3/2} e^{-\frac{\pi(M_0^2-2H^2)}{|g\dot{\phi}|}} \left(\frac{a_*}{a_0}\right)^3H^3\times (4\pi\eta_0^2\Delta\eta_{\rm rec})\\
& = \frac{1}{2\pi^2}\left(\frac{g\dot{\phi}}{H_*^2}\right)^{3/2} e^{-\frac{\pi(M_0^2-2H^2)}{|g\dot{\phi}|}} \left(k_*\eta_0\right)^3\times \frac{\Delta\eta_{\rm rec}}{\eta_0},
\end{align}
where $a_*$ is the scale factor at the time of particle production. In the last relation, we have used the fact that a mode with comoving momentum $k$ crosses the horizon at time $t$ when $k = a(t)H_*(t)$. Thus, $k_{*}=a_*H_*$ is the mode that exits the horizon at the time of particle production.  The ratio involving the thickness of the surface of recombination is given by, $\Delta \eta_{\rm rec}/\eta_{0}\approx 10^{-3}$ with $\Delta \eta_{\rm rec}\approx 19$~Mpc~\cite{Hadzhiyska:2018mwh}. Therefore, the number of hotspots inside the surface of last scattering depends on our model parameters $g,M_0$ and $\eta_*$ as,
\begin{align}\label{eq.nhotspotsfinal}
N_{\rm HS}\simeq 1.3\times 10^{8}\left(\frac{g}{3}\right)^{3/2} e^{-\frac{\pi}{g} \left(\frac{M_0}{60H}\right)^2} \left(\frac{100~\text{Mpc}}{\eta_*}\right)^3.
\end{align}
Here we have used the horizon crossing relation $k_*\eta_*=1$. The exponential sensitivity to $M_0$ comes from the non-adiabaticity factor $e^{-\frac{\omega^2}{\dot{\omega}}}$ mentioned above, but we see that even for $M_0 \gg H_*$, there can still be an appreciable number of hotspot production.  The scaling $\propto1/\eta_*^3$ denotes the volume dilution of hotspots after their production. Finally, $g$ controls the number of hotspots and, as we will see below, the temperature of individual hotspots as well.

\subsection{Backreaction constraints}\label{sec.back}
Before exploring PHS phenomenology and detectability in more detail, we review the existing constraints on the setup. First we want to ensure that the production of heavy particles does not affect the background inflating spacetime~\cite{Chung:1998bt,Flauger:2016idt}. The EOM obeyed by the inflaton reads as,
\begin{align}
    \ddot{\phi}+3H\dot{\phi}+\frac{\partial V}{\partial \phi}=0.
\end{align}
In the presence of the interaction implied by the $\phi-$dependent mass term in Eq.~\eqref{eq.heavylag} we get
\begin{align}
\frac{\partial V}{\partial \phi} = \frac{\partial V_\phi}{\partial \phi}+g(g\phi-\mu)\langle\sigma^2\rangle,
\end{align}
where $V_\phi$ is the background inflationary potential that drives the inflationary expansion and $\langle \sigma^2\rangle$ is the vacuum expectation value with respect to the early time vacuum. To ensure the usual slow-roll dynamics for the inflaton, $3H\dot{\phi}\approx -\frac{\partial V_\phi}{\partial \phi}$,
we demand
\begin{align}
g(g\phi-\mu)\langle\sigma^2\rangle \ll \frac{\partial V_\phi}{\partial \phi}\sim H_*\dot{\phi}.
\end{align}
Using Eq.~\eqref{timedepmass} we can rewrite the above condition as,
\begin{align}
g^2\dot{\phi}|\ln(\eta/\eta_*)|\langle\sigma^2\rangle \ll H_*^2\dot{\phi}.
\end{align}
For $M\sim g\dot{\phi}|\ln(\eta/\eta_*)|/H_*$, this implies
\begin{align}
&g M\langle\sigma^2\rangle \ll H_*\dot{\phi} \nonumber\\
\Rightarrow &\frac{g}{8\pi^3}\left(\frac{g\dot{\phi}}{H_*^2}\right)^{3/2} e^{-\frac{\pi(M_0^2-2H^2)}{|g\dot{\phi}|}} \ll \frac{\dot{\phi}}{H_*^2}\approx 60^2 \label{eq.velconstraint},
\end{align}
where we have used the fact that $n\sim M\langle\sigma^2\rangle$ is the number density of the produced $\sigma$ particles around $\eta_*$, given by eq.~\eqref{eq.nodensity}.

\begin{figure}
\center{
\includegraphics[width=10 cm]{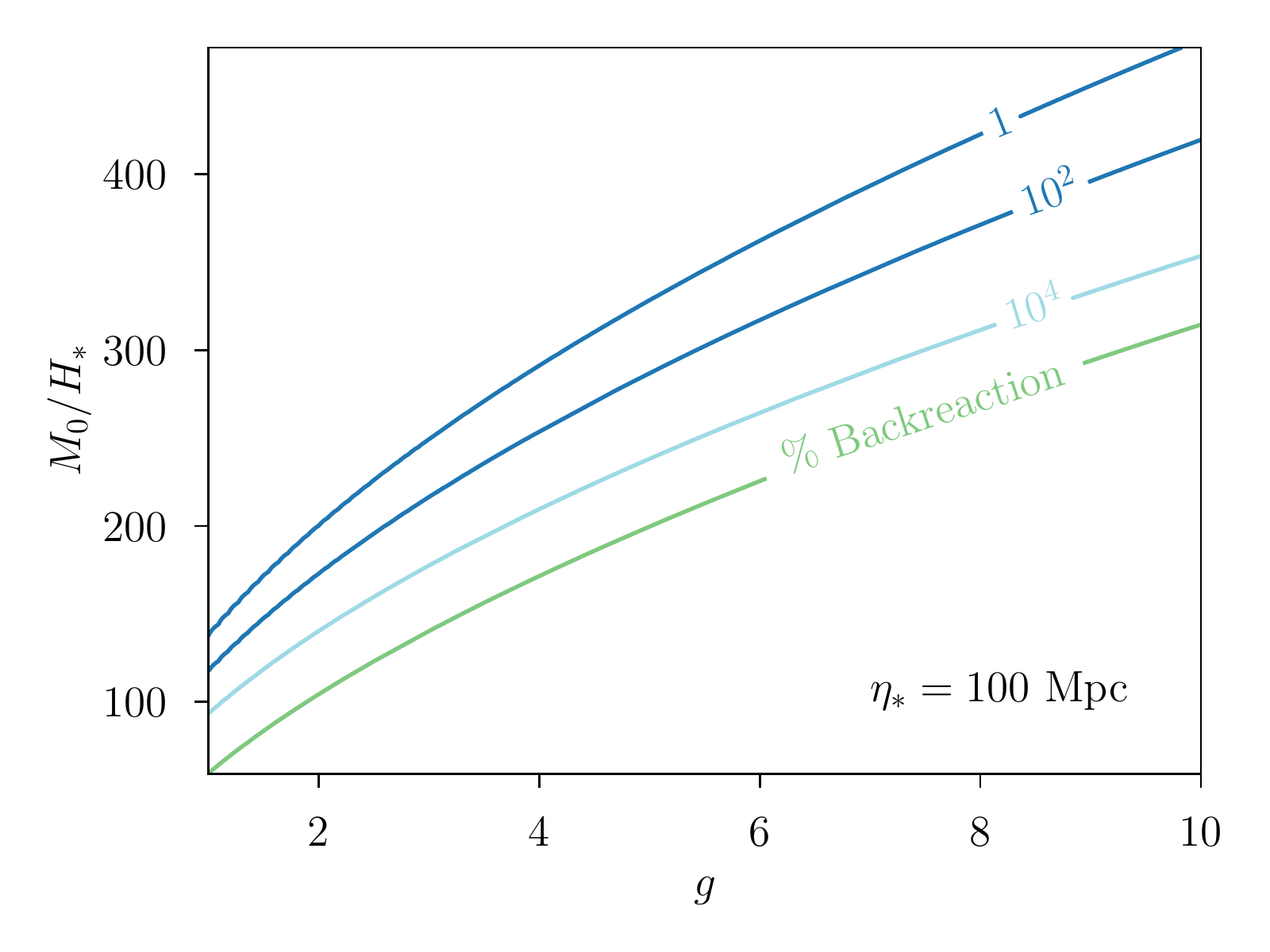}
}
\caption{Allowed parameter space for subdominant backreaction on the homogeneous inflationary dynamics for $\eta_*=100$~Mpc. Below the contour labeled ``\% Backreaction'', the backreaction of produced particles on the slow-roll dynamics is more than a percent around a scale $k\sim 1/\eta_*$. For our parameter choices, we stay above this line. The other contours for different values of $N_{\rm HS}$ (see Eq.~\eqref{eq.nhotspotsfinal}) show that an appreciable number of hotspot can still be produced with negligible backreaction.}\label{fig.paramspace}
\end{figure}

To ensure that particle production does not deplete the kinetic energy of inflaton, we require the energy density of the produced $\sigma$ particles to be smaller than $\dot{\phi}^2$, i.e.,
\begin{align}
\rho_\sigma \sim M n \ll \dot{\phi}^2.
\end{align}
Using Eq.~\eqref{timedepmass} once again, we can rewrite the above constraint as,
\begin{align}
\label{eq:constraint2}
\frac{g}{8\pi^3}\left(\frac{g\dot{\phi}}{H_*^2}\right)^{3/2} e^{-\frac{\pi(M_0^2-2H^2)}{|g\dot{\phi}|}}  |\ln(\eta/\eta_*)| \ll \frac{\dot{\phi}}{H_*^2},
\end{align}
which for $\ln(\eta/\eta_*)\sim 1$ reduces to constraint~\eqref{eq.velconstraint}. The constraint in relation~\eqref{eq.velconstraint} is shown in Fig.~\ref{fig.paramspace} where the left hand side is a percent of the right hand side and above the green line labeled ``\% Backreaction'' the backreaction will be more than a percent. Various contours of $N_{\rm HS}$ are also shown which determine how many hotspots can be produced with negligible backreaction on the inflationary dynamics. The constraints should be thought of as conservative, as the left hand sides of Eqs.~\eqref{eq.velconstraint}, \eqref{eq:constraint2} involve the number density at $|\eta_*|$, the maximal value, and have no dilution effects. As presented, the constraints are primarily on $g$ and $M_0/H_*$, with only a very mild dependence on $\eta_*$. 


\section{Hotspot phenomenology}\label{sec.hspheno}

Having estimated the strength of particle production, now we focus on the effect of the produced particles on curvature perturbation and, eventually, on the CMB. We do this in two steps. First, we compute the curvature perturbation due to a massive particle. Then, we include the effects of sub-horizon transfer function to see the final form of the temperature anisotropy.

\subsection{Curvature perturbation due to a massive particle}
To compute the curvature perturbations due to $\sigma$ production, we employ a sudden transition approximation, i.e., we assume that the particles become non-relativistic immediately after their production. This is justified through Eq.~\eqref{eq.beta2}, which shows that particle production is exponentially suppressed unless physical momenta of the particles $k\eta_* H_*\lesssim M_0$, for $M_0^2\gtrsim g\dot{\phi}$ (as we will choose later). Furthermore, after their production the particle masses increase, whereas the physical momenta redshift further. Therefore, the action due to a single heavy particle can be written in a non-relativistic form as,
\begin{equation}\label{eq.actionparticle}
S_{\rm particle}=-\int d\tau\, M(t,\vec{x})\sqrt{-\dot{x}^{\mu}(\tau)^2}\approx-\int dt \sqrt{-g_{00}}\,M(t)\,,
\end{equation}
where $x^{\mu}=(t,\vec{0})$ is the location of the static massive particle, chosen to be at the origin. To see how this contributes to the curvature perturbation, we parametrize the metric fluctuations as~\cite{Arnowitt:1962hi}, 
\begin{equation}
{\rm d}s^2=-N^2{\rm d}t^2+h_{ij}({\rm d}x^i+N^i{\rm d}t)({\rm d}x^j+N^j{\rm d}t)\,,\quad h_{ij}=e^{2Ht}\left[(1-2\psi)\delta_{ij}+\gamma_{ij}\right]\,,
\end{equation}
with the condition $\partial_i\gamma_{ij}=0$ and $\gamma_{ii}=0$. In this gauge there is no inflaton fluctuation, $\delta\phi=0$. The quantities $N$ and $N^i$ have vanishing time derivative and therefore are not dynamical. In particular, $N$ obeys the algebraic relation~\cite{Maldacena:2002vr},
\begin{equation}
N=1-\frac{\dot {\psi}}{H_*}=1+\frac{\dot {\zeta}}{H_*}.
\end{equation}
In the last line we have expressed $N$ in terms of the comoving curvature perturbation $\zeta$~\cite{Malik:2008im},
\begin{equation}
\zeta\equiv-\psi-H_*\frac{\delta\phi}{\dot\phi}=-\psi,
\end{equation}
as appropriate in the $\delta\phi=0$ gauge. We can now rewrite the action integral~\eqref{eq.actionparticle} into
\begin{equation}
S_{\rm particle}=-\int dt\sqrt{-g_{00}}M=-\int dt\, N\,M=-\int dt\left(1+\frac{\dot \zeta}{H_*}\right)\,M=-\int dt\,M-\int dt\,\dot{\zeta}\,\frac{M}{H_*}\,.
\end{equation}
The last integral determines the effect of a heavy particle on the curvature perturbation. In terms of the conformal time $\eta$, the $\zeta$-tadpole term is 
\begin{equation}\label{eq:tadpole}
S_{\rm particle}\supset-\int d\eta\,\partial_{\eta}\zeta\,\frac{M(\eta)}{H_*}\,.
\end{equation}
The above interaction term is non-trivial due to the time dependence of the mass $M(\eta)$, and it gives rise to a non-zero one point function of $\zeta$, $\langle\zeta_{\rm HS}\rangle$. We will see below that after including the effects of transfer function, $\langle\zeta_{\rm HS}\rangle$ above would give rise to hotspots (HS) on the CMB for our parameter choices, as opposed to cold spots. This one-point function $\langle \zeta_{\rm HS}\rangle$ can be calculated using the ``in-in'' formalism along the lines of~\cite{Maldacena:2015bha} (for a derivation see Appendix~\ref{app:profile}), 
\begin{equation}\label{eq:hsprofile}
\langle \zeta_{\rm HS}(r)\rangle=\frac{H_*}{8\pi\epsilon M_{\rm pl}^2}\begin{cases} M(\eta=-r), & \mbox{if } |\eta|\leq|\eta_*|\\ 0, & \mbox{if } |\eta|>|\eta_*| \end{cases}\,,
\end{equation}
where $\epsilon=-\dot{H}_*/H_*^2$. 

We can understand the above result intuitively as follows. Due to causality, the massive particle can not change the curvature perturbation on scales larger than the horizon size at the time of particle production, determined by $\eta_*$. Hence there is no effect for $|\eta|>|\eta_*|$.\footnote{Strictly speaking, particle production also happens for times slightly earlier than $\eta_*$ and hence, $\langle \zeta_{\rm HS}(r)\rangle$ would be non-zero for $|\eta|$ slightly bigger than $|\eta_*|$. However, already for $|\eta|\sim 1.5|\eta_*|$, it can be checked that the effective mass is too large for particle production to happen for our parameter choices.} After production, the particle mass keeps changing gradually due to the slow-roll of the inflaton field and this temporal change is recorded through a spatial variation of curvature. For example,
a spherically symmetric shell that exits the horizon at $|\eta_1|<|\eta_*|$ is perturbed according to $M(\eta_1)$. Another shell that exits at a later time $|\eta_2|<|\eta_1|$ is perturbed according to $M(\eta_2)$. Thus a hotspot shows up with a comoving size $|\eta_*|$, and the radial curvature profile inside that region is determined by the time evolution $M(\eta)$ during inflation and after $\eta_*$.

For our model, $M(\eta)$ varies as in Eq.~\eqref{timedepmass}. Therefore, for $r\lesssim \eta_*$ the hotspot profile can be rewritten after dropping the subleading contributions due to $M_0$ and using $\dot{\phi}_0=\sqrt{2\epsilon}H_* \mpl$,
\begin{equation}\label{eq.Rofr}
\langle \zeta_{\rm HS}(r)\rangle\approx\left[\frac{g}{2}\log\left(\frac{\eta_*}{r}\right)\right]\frac{H_*}{2\pi\sqrt{2\epsilon} \mpl}\theta(\eta_*-r)\,.
\end{equation}
Given the size of the standard inflationary quantum fluctuation at any point, $\langle\zeta_{\rm q}^2\rangle^{\frac{1}{2}}\simeq \frac{H_*}{2\pi\sqrt{2\epsilon} \mpl}$, the visibility of the hotspot over such fluctuations
\begin{align}\label{eq.coverq}
    \frac{\langle \zeta_{\rm HS}(r)\rangle}{\langle\zeta_{\rm q}^2(r)\rangle^{\frac{1}{2}}} \simeq \frac{g}{2}\log\left(\frac{\eta_*}{r}\right),
\end{align}
is mostly controlled by the coupling $g$. The above relation also indicates that for visibility of the hotspot signal, we typically need $g\gtrsim 1$, as will also be seen below through our numerical simulations. Given the above relation, we see that while $M_0$ determines the particle production rate, the hotspot perturbation is determined by the time-dependent mass \textit{away} from the minimum. This is so because already after one e-folding, $M(\eta)$ gets dominated by the first term in Eq.~\eqref{timedepmass}. Therefore, the hotspot profile for $r<\eta_*/\text{few}$ is determined by $g$.

\subsection{Temperature anisotropy from curvature perturbation}
Having calculated the curvature perturbation due to the massive particle, we now connect it to the observed temperature fluctuation on the sky. We write the metric fluctuations in the Newtonian gauge as,
\begin{align}
ds^2 = -(1+2\Psi)dt^2+a^2(t)(1+2\Phi)\delta_{ij}dx^{i}dx^{j}.    
\end{align}
We can write the temperature fluctuation on the sky corresponding to a Fourier mode $\vec{k}$ as, $\delta T(\vec{k},\hat{n},\eta_0)/T\equiv\theta(\vec{k},\hat{n},\eta_0)$, where $\hat{n}$ denotes the direction on the sky, $\eta_0$ is the present (conformal) age and $T$ is the average CMB temperature. Then we can rewrite this in terms of Legendre polynomials as,
\begin{align}
\theta(\vec{k},\hat{n},\eta_0) = \sum_{l}i^l (2l+1)\mathcal{P}_l(\hat{k}\cdot\hat{n})\theta_l(k,\eta_0), 
\end{align}
where the moments $\theta_l(k,\eta_0)$ get contribution from Sachs-Wolfe (SW), integrated Sachs-Wolfe (ISW) and Doppler effects. Given the spherically symmetric profile of the hotspot, the Doppler contribution to $\theta_l(k,\eta_0)$ is very small, as can be checked explicitly. Therefore, we focus on the SW and the ISW contributions which are given by,
\begin{align}
\theta_l(k,\eta_0)&\approx (\theta_0(k,\eta_{\rm rec})+\Psi(k,\eta_{\rm rec}))j_l(k(\eta_0-\eta_{\rm rec}))+\int_{0}^{\eta_0}d\eta e^{-\tau} \left(\Psi'(k,\eta)-\Phi'(k,\eta)\right)j_l(k(\eta_0-\eta))\nonumber,\\
&\equiv f_{\rm SW}(k) \langle\zeta(\vec{k})\rangle + f_{\rm ISW}(k) \langle\zeta(\vec{k})\rangle,
\label{eq:subhorizon}
\end{align}
where $j_l(x)$ is the spherical Bessel function and $\eta_{\rm rec}$ denotes the time of recombination. We have also isolated the two functions $f_{\rm SW}(k)$ and $f_{\rm ISW}(k)$ to characterize the SW and ISW contributions after mode re-entry with $\zeta(\vec{k})$ denoting the primordial perturbation as before. In practice, we extract $f_{\rm SW}(k)$ by running \texttt{CLASS} for a purely $\Lambda\text{CDM}$ setup. We numerically fold in the spherical Bessel function with the \texttt{CLASS} output of $\theta_0(k,\eta), \Psi(k,\eta)$ perturbations 
from which the primordial fluctuation $\langle \zeta(\vec k)\rangle$ is already factored out. This leaves us with $f_{\rm SW}(k)$.
Similarly, numerically integrating $\Psi'(k,\eta), \Phi'(k,\eta)$ with the spherical Bessel function and the optical depth (second expression in Eq.~\eqref{eq:subhorizon}) 
gets us $f_{\rm ISW}(k)$. The $f_{\rm SW}(k), f_{\rm ISW}(k)$ extracted in this fashion can be applied to $\langle \zeta_{\rm HS}(k)\rangle$ as subhorizon physics is independent of the origin of the initial curvature perturbation.

With $f_{\rm SW}(k)$ and $f_{\rm ISW}(k)$ in hand, we are now ready to compute the temperature perturbation due to the heavy particle. We rewrite the momentum space expression of $\langle\zeta_{\rm HS}\rangle$ derived in Eq.~\eqref{eq.zetak} as,
\begin{align}\label{eq.zetaHS}
\langle\zeta_{\rm HS}(\vec{k})\rangle = e^{-i\vec{k}\cdot \vec{x}_{\rm HS}}\frac{f(k\eta_*)}{k^3},    
\end{align}
where the function $f(x)=\frac{gH^2}{\dot{\phi}_0}\left(\text{Si}(x)-\sin(x)\right)$. Thus we can write the temperature fluctuation at our location $\vec{x}_0$ as,
\begin{align}
\theta(\vec{x}_0,\hat{n},\eta_0)=\int \frac{d^3\vec{k}}{(2\pi)^3}e^{i\vec{k}\cdot(\vec{x}_0-\vec{x}_{\rm HS})}\sum_{l}i^l (2l+1) \mathcal{P}_l(\hat{k}\cdot\hat{n})\left(f_{\rm SW}(k)+f_{\rm ISW}(k)\right)\frac{f(k\eta_*)}{k^3}.    
\end{align}
Using the plane wave expansion,
\begin{align}
e^{i\vec{k}\cdot\vec{r}}=\sum_{\ell=0}^\infty i^l (2l+1) j_l(kr) \mathcal{P}_l(\hat{k}\cdot\hat{r}),   
\end{align}
the distance to the hotspots,
\begin{align}
\vec{x}_0-\vec{x}_{\rm HS} \approx -(\eta_0-\eta_{\rm rec})\hat{n}_{\rm HS},
\end{align}
and the identity between Legendre polynomials and spherical harmonics $Y_{lm}$,
\begin{align}
\mathcal{P}_l(\hat{k}\cdot\hat{n})=\frac{4\pi}{(2l+1)}\sum_{m=-l}^{l}Y_{lm}(\hat{n})Y_{lm}^*(\hat{k}),    
\end{align}
we finally get\footnote{Note that, Eqs.~\eqref{eq:subhorizon}-\eqref{eq.thprofile} should technically be carried out using the total (hotspot + inflaton) curvature perturbation as input. However, because the inflaton contribution is isotropic (no $\exp({i\vec{k}\cdot\vec x_{\rm HS}})$), its contribution to $\theta_0$ vanishes.}
\begin{align}\label{eq.thprofile}
\theta(\vec{x}_0,\hat{n},\eta_0) = \frac{4\pi}{(2\pi)^3} \int_0^\infty \frac{dk}{k}\sum_{l}j_{l}(k\eta_0-k\eta_{\rm rec})(2l+1)\mathcal{P}_l(\hat{n}\cdot\hat{n}_{\rm HS})\left(f_{\rm SW}(k)+f_{\rm ISW}(k)\right) f(k\eta_*).   
\end{align}
Eq.~\eqref{eq.thprofile} gives the temperature profile around the hotspot as a function of the angle between the direction of observation $\hat{n}$ and the direction of the hotspot $\hat{n}_{\rm HS}$. We will use this profile in the subsequent analysis of hotspots in CMB simulations.

\begin{figure}[h]
    \centering
    \includegraphics[width=13cm]{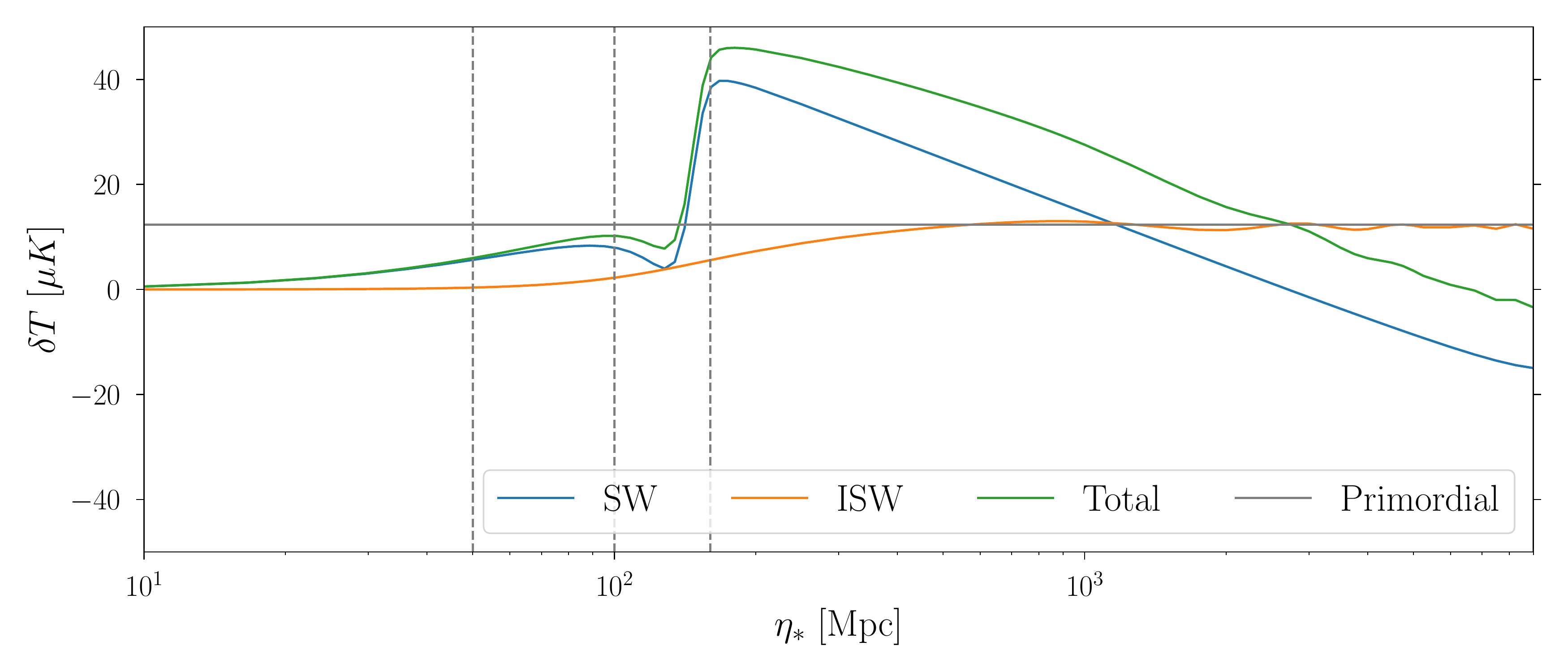}
    \caption{Temperature of the central pixel (green) of a hotspot originating from a heavy particle for $g=1$, based on Eq.~\eqref{eq.thprofile} with $\hat{n}=\hat{n}_{\rm HS}$. This is done by summing the SW (blue) and the ISW (orange) effects, with the Doppler effect being subdominant. The green line shows how the observed anisotropy varies as a function of ``size'' of the hotspot, characterized by the size of the comoving horizon at the time of particle production, $\eta_*$. The gray line shows the going rate of the temperature fluctuation due to only the standard primordial quantum fluctuations (i.e., without any subhorizon effect), determined by $\frac{1}{5}\times\frac{g}{2}\langle\zeta_{\rm q}^2\rangle^{\frac{1}{2}}\times 2.7$~K. Here $\langle\zeta_{\rm q}^2\rangle^{\frac{1}{2}}$ is given below Eq.~\eqref{eq.Rofr} and the factor $1/5$ characterizes a flat superhorizon transfer function. The dashed vertical gray lines show the benchmark choices for the hotspot size chosen in the subsequent discussion. For $\eta_*\gtrsim$~6600~Mpc, the central temperatures associated with the heavy particles become negative since the (now negative) SW effect dominates over the (positive) ISW effect, and consequently, they give rise to \textit{cold}spots as opposed to \textit{hot}spots.}
    \label{fig:transfer}
\end{figure}

To obtain the central temperature of a hotspot we can set $\hat{n}\approx\hat{n}_{\rm HS}$ implying  $\mathcal{P}_l(\hat{n}\cdot\hat{n}_{\rm HS})\approx 1$. For small enough $\eta_*$, the ISW contribution can be neglected. In that case, a simple expression of the central temperature can be obtained by noting $f_{\rm SW}(k)=T_{\rm SW}(k)j_l(k\eta_0-k\eta_{\rm rec})$ with $T_{\rm SW}(k)$ being the SW transfer function (after factoring out $\zeta(\vec{k})$). Upon further using the approximation $\sum_{l}j_{l}^2(x)(2l+1)\approx 1$, we get 
\begin{align}
\theta(\vec{x}_0,\hat{n}_{\rm HS},\eta_0) \approx \frac{4\pi}{(2\pi)^3} \int_0^\infty \frac{dk}{k}T_{\rm SW}(k) f(k\eta_*)~~\text{for small}~~\eta_*.   
\end{align}
For larger $\eta_*$, we need to keep the ISW contribution as well. The full result for the central temperature for various values of $\eta_*$ is shown in Fig.~\ref{fig:transfer}. We see the net effect of the subhorizon evolution is to give rise to a \textit{hot}spot, as opposed to a \textit{cold}spot. In Fig.~\ref{fig:hsprofile} we will show the angular profile for a given hotspot as a function of the angular distance from the center of the hotspot.

\subsection{Separation between hotspots}
\label{sec.sep}

In the above, we have seen how a single heavy particle gives rise to a hotspot in curvature perturbation. However, due to momentum conservation, heavy particles are produced in pairs. Hence as long as the heavy particles are sufficiently separated, what we expect to see is not a single hotspot but rather a pair of them.

To get an idea about the separation between the two members of a hotspot pair, we recall from Eq.~\eqref{eq.beta2} that the particles are produced either semi- or non-relativistically. Furthermore, within one Hubble time after particle production, the effective mass of a particle increases from $\approx M_0$ to $\frac{g\dot{\phi}}{H_*}|\ln(\eta/\eta_*)|$, as seen from Eq.~\eqref{timedepmass}. Hence, most of the particle propagation due to non-zero physical momentum $k_{\rm phys}= ka^{-1}\sim (g\dot\phi)^{1/2}$ happens immediately after production. The resulting comoving distance travelled by a massive particle up to conformal time $\eta_f$ can be estimated as,
\begin{align}\label{eq.sep}
\int_{t_{*}}^{t_f}\frac{dt}{a}\frac{k_{\rm phys}}{M}=\int_{\eta_{*}}^{\eta_f} d\eta \frac{k|\eta| H_*}{M} \sim \frac{k\eta_*^2H}{M_0}\lesssim |\eta_*|.    
\end{align}
To get the above we have used $M\approx M_0$ at production and $k_{\rm phys}\approx k\eta_*H_*\lesssim M_0$.
Thus we get the intuitive result that for semi-relativistic production, the particles can travel a horizon-sized distance in one Hubble time, whereas for non-relativistic production, the particles mostly remain where they were produced. Given this, we will model the separation between the particles of the same pair to be a random uniform distribution between $0$ and $\eta_*$.

\subsection{Effect of localized hotspots on the CMB power spectrum}\label{sec:ClTT}
Before discussing in detail how to search for the above-mentioned hotspots, we can ask whether the CMB power spectrum is already good enough for that purpose. To get an analytical understanding, we consider a toy model in which a hotspot is represented by a delta function on the CMB sky. Therefore, the CMB temperature fluctuation reads as
\begin{align}
    \Delta T(\theta,\phi)= \Delta T_{\rm{inf}}(\theta,\phi)+T_h \delta(\theta-\theta_*)\delta(\phi-\phi_*).
\end{align}
Here, $\Delta T_{\rm{inf}}$ denotes the standard inflaton-sourced CMB fluctuation and $(\theta_*,\phi_*)$ is the location of the hotspot with a temperature $T_h$ ($\propto\, g$). By doing the standard spherical harmonics decomposition
\begin{align}
\Delta T(\theta,\phi) = \sum_{l,m}a_{lm}Y_{lm}(\theta,\phi),
\end{align}
we get,
\begin{align}
    a_{lm}=a_{lm,\rm inf}+T_h Y^*_{lm}(\theta_*,\phi_*)\sin(\theta_*).
\end{align}
Therefore, the temperature power spectrum estimator $C_\ell^{\rm TT}$
reads as,
\begin{align}
C^{\rm TT}_\ell = \frac{1}{(2l+1)}\sum_{m=-l}^{l} |a_{lm}|^2 = C_{\ell,\rm inf}+\frac{T_h^2}{4\pi}\sin^2(\theta_*),   
\end{align}
where the interference term vanishes between the two uncorrelated perturbations. Conventionally, one looks at $\mathcal{D}_{\ell}\equiv \ell(\ell+1)C_\ell/(2\pi)$ rather than $C^{\rm TT}_\ell$ for the power spectrum:
\begin{align}\label{eq.DlTT}
\mathcal{D}^{\rm TT}_\ell = \mathcal{D}_{\ell,\rm inf}+\frac{\ell(\ell+1)}{8\pi^2}T_h^2\sin^2(\theta_*).    \end{align}
Including the $g$ dependence in $T_h$ and multiplying by the number of hotspots yields a PHS contribution to $\mathcal{D}^{\rm TT}_\ell \sim g^2\, N_{\rm HS}\, \ell^2$ that increases at higher $\ell$-modes. However, a more realistic model of the hotspot must include their finite profile size. For hotspots with size $\sim \eta_*$, $\ell$-modes with $\ell\gg \sqrt{4\pi}\eta_*^{-1}$ correspond to epochs after the particles were produced and should make negligible contribution to $\mathcal{D}^{\rm TT}_\ell$. Production of modes with $k\gg 1/\eta_*$ is also Boltzmann suppressed as seen from Eq.~\eqref{eq.beta2}. Combining the growth at low $\ell$ from the toy model with vanishing effect from large $\ell$, we expect the PHS impact on $\mathcal{D}^{\rm TT}_\ell$ to be most prominent at $\ell_*$. Depending $\ell_*$, and $g$, such an effect could be hidden within the uncertainties on the power spectrum. So, while the power spectrum will constrain the PHS model to some degree, it is worthwhile to consider other detection possibilities as well.

\section{Detection strategy}\label{sec.detstrat}
Here we discuss the possibility of identifying PHS using a simplified search for the hotspot signals in position space. We begin with some comments about the various backgrounds to the PHS signal and then motivate two different search strategies.

There are three types of backgrounds to consider when searching for PHS: i.) detector noise, ii.) the astrophysical foreground,  and iii.) background from the standard primordial fluctuations. Detector noise depends on the details of the instrument and can be reduced further at future CMB experiments. The foreground emissions arising from compact objects such as galaxies, galaxy clusters, gas, and dust that exist between the last scattering surface and us, can also produce localized signals. Since the compact sources usually produce higher frequency signals than the CMB photons, the comparison between different frequency maps has been a powerful tool to subtract the foreground in CMB studies~\cite{Planck:2013qym}. Therefore, it is possible to remove the foregrounds by comparing the sky maps taken at different frequency bands. Moreover, the shape of the compact objects are usually not spherically symmetric and may be distinguished from the signal hotspots that we are looking for. Therefore, in the following, we only consider the background from the primordial, almost Gaussian fluctuations when studying the PHS signal. We will assume the CMB maps are masked to reduce the astrophysical foregrounds and badly-conditioned pixels and retain only $60\%$ of the sky for the analysis. The number is similar to the sky fraction used in the Planck analysis~\cite{Aghanim:2018eyx}. 

We consider two types of signature from the PHS.
\begin{itemize}
\item[1.)] {\bf Position-space search:} The PHS show up as localized objects in the sky, motivating a search in position space. Even in idealized scenarios, the primordial quantum fluctuation can still generate a background for the PHS and, unlike the astrophysical foregrounds, background signals from the primordial fluctuation follow the same blackbody distribution as the PHS signal. We need to rely on the hotspots' detailed properties, such as their temperature and distribution, to distinguish them from the background. We therefore consider a ``cut-and-count'' analysis in position space by applying temperature cuts on the entire sky to veto the standard CMB fluctuations.  While past position space searches have utilized  wavelet analysis based approach~\cite{Gonzalez-Nuevo:2006ood,Lopez-Caniego:2006kci,Sanz:1999xu} or a more dedicated template fit (e.g.,~\cite{Osborne:2013jea}), in this work we explore how much an even simpler ``cut-and-count'' method can be useful.
    \item[2.)] {\bf Power spectrum analysis:} As discussed in Sec.~\ref{sec:ClTT}, the presence of PHS also modifies correlation functions of CMB fluctuations. The existing measurements of the CMB power spectrum, the searches for bispectrum and trispectrum can constrain the number of PHS. Our primary concern will be the corrections to the power spectrum $C_{\ell}^{\rm TT}$, though we will also briefly mention the rough (scale-dependent) amplitude of the non-Gaussianity parameter $f_{\rm NL}$ for the same number of PHS under consideration. There will also be corrections to power spectra $C_{\ell}^{\rm TE,EE}$ involving $E$-mode polarization. Since such corrections are of similar order-of-magnitude compared to $C_{\ell}^{\rm TT}$, to get a first bound on our parameter space, we will not consider polarization spectra in this work.
\end{itemize} 

Before discussing these two search strategies in detail, we step through the numerical simulation recipe we follow to generate the PHS signal and mock CMB maps.

\subsection{Simulating the PHS signal}
In this section, we describe a detailed methodology for simulating CMB maps and PHS signals. The first step is to generate a temperature power spectrum $C_{\ell}^{\rm TT}$ using the Boltzmann code \texttt{CLASS v2.9}~\cite{Lesgourgues:2011re,Blas:2011rf} with the best fit $\Lambda$CDM  parameters
\begin{eqnarray} \label{eq.default}
\{\omega_{\rm cdm},\omega_b, h,  A_s, n_s, \tau_{\rm reio}\}=\{0.120, 0.022, 0.676,  2.22 \times 10^{-9} , 0.962, 0.0925  \}  \;.
\end{eqnarray}
The next step is to convert this angular power spectrum into sample CMB maps using \texttt{HEALPix}~\cite{Gorski:2004by}. A key input for the \texttt{HEALPix} simulation is $N_{\rm side}$, which controls the resolution of the CMB maps: $N_{\rm pixels} = 12\, N^2_{\rm side}$. The number of pixels can be compared to the number of independently fluctuating temperature patches $\sum_{\ell=1}^{\ell_{\rm max}}(2\ell+1)$. Thus a simulation using $N_{\rm side} = 256$ corresponds to an $\ell_{\rm max} \simeq 880$ while $N_{\rm side} = 2048$ corresponds to $\ell_{\rm max} \simeq 7000$.\footnote{Per \texttt{HEALPix} convention, $N_{\rm side}$ is taken to be a power of 2.} The price for larger $N_{\rm side}$ is significantly slower simulation speed. Balancing resolution and computational cost, we proceed with a hybrid approach. Specifically, for the cut-and-count analysis, which requires a large number of simulated CMB images, we use lower resolution, typically $N_{\rm side}=256$ or $512$. However, for the power spectrum analysis we do not need to generate multiple images, but we want to capture the impact of the PHS up to $\ell \sim 2000$ (Planck 2018 range) -- so we use $N_{\rm side}=2048$. A few other useful relations to keep in mind are i.) how to convert between $\ell$ mode and angle on the sky: $\theta = \sqrt{4\pi}\ell^{-1}_{\rm max}$ , and ii.) how to convert between $\eta$ and  $\ell$-mode : $\ell_* \simeq \eta_0/\eta_*$, with $\eta_0\approx13.8$ Gpc being the current conformal age of the Universe.

\begin{figure}
    \centering
    \includegraphics[width=9cm]{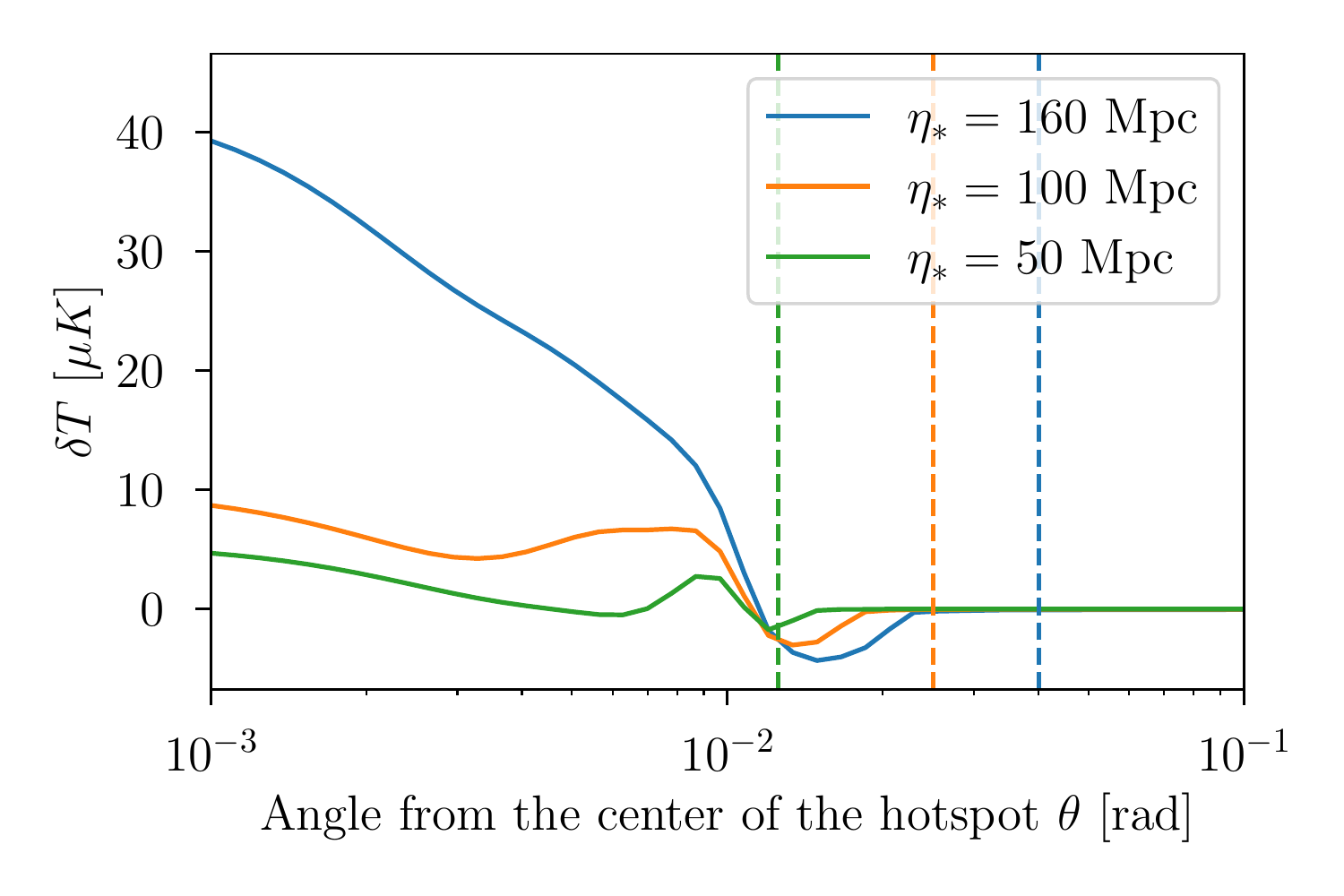}
    \caption{Angular profile of a hotspot for three different choices $\eta_*=160,100,50$~Mpc based on Eq.~\eqref{eq.thprofile}. The vertical lines denote the angular size of the corresponding hotspots and is related to $\eta_*$ by, $\theta_*=\sqrt{4\pi}\eta_*/\eta_0$. Here we have set $g=1$.}
    \label{fig:hsprofile}
\end{figure}

For a given \texttt{HEALPix} CMB map, we next add a population of PHS signals based on the profiles that we derived above in Eq.~\eqref{eq.thprofile}. From Eqs.~\eqref{eq.zetaHS},~\eqref{eq.sep}, we see that the angular size, separation and temperature of the PHS signal are controlled by the heavy scalar-inflaton coupling $g$ and comoving horizon size $\eta_*$ at particle production, while the overall number of PHS also depends on $M_0$. For our simulations, we pick some benchmark values of $\eta_*$ and $g$, then use the signal and background analysis to set limits on $M_0$. The benchmark parameter points we choose are
\begin{equation}\label{eq.sigl}
\eta_*=160,\, 100,\, 50\,{\rm Mpc}\,,\quad g = 2, 3, 6, 10.
\end{equation}
These $\eta_*$ values roughly correspond to 
$\ell_*\approx 85,\,140,\,280$. The shape of the temperature profiles (i.e., setting $g=1$ in Eq.~\eqref{eq.thprofile}) is controlled by $\eta_*$ alone and is shown in Fig.~\ref{fig:hsprofile} as a function of the angle $\cos\theta=\hat n\cdot\hat{n}_{\rm HS}$ away from the center of the hotspot. The vertical lines in the figure show the angles $\sqrt{4\pi}/\ell_*$ that correspond to the angular sizes of $\eta_*$ in the sky.
We pick these benchmark values to show examples with  different corrections due to the transfer function. If $\eta_*$ is much smaller than $50$~Mpc, suppression from the transfer function will diminish the signal to be below the primordial fluctuation even for a large coupling $g$. We start the maximum horizon from $160$~Mpc so that we can compare the result of the position space search to the $C_{\ell}^{\rm TT}$ analysis with a reasonably small uncertainty from cosmic variance.

For the $\eta_*=160$~Mpc benchmark, we use $N_{\rm side}=256$ maps, corresponding to $\ell_{max} \simeq 880$ or angular resolution of $\approx\sqrt{4\pi}/\ell_{\rm max}=0.004$. The $\delta T$ profile in this case effectively extends to $\theta\approx 0.02$, as is shown in Fig.~\ref{fig:hsprofile}, therefore we divide the hotspot radius into $5.5$ pixels in the \texttt{HEALPix} image, where the half pixel corresponds to the central pixel of the hotspot, so that in the 2D pixel-space, an individual hotspot spans a $11\times 11$ block. The temperature in each hotspot pixel is set to the average value of $\delta T$ across that pixel, following profiles in Fig.~\ref{fig:hsprofile} and scaled by the approrpriate benchmark value of $g$. After adding one hotspot for the PHS signal, we add its partner -- a second, identical spot, but with its center offset from the original spot. The offset is determined via a uniform random distribution 
up to a maximum angle of $\approx\sqrt{4\pi}/\ell_*$ (shown as vertical lines in Fig.~\ref{fig:hsprofile}). For $\eta_*=160$~Mpc, the maximum separation corresponds to 10 pixels ($\Delta\theta\approx 0.04$).

The PHS signal generation procedure for the  $\eta_*=100$~Mpc and  $\eta_*=50$~Mpc benchmarks is identical. For $\eta_*=100$~Mpc, the $\delta T$ profile effectively extends to $\theta\approx 0.02$. We again use $N_{\rm side} = 256$, so the hotspots have a radius of $5.5$ pixels and a maximum separation of $6$ pixels. For $\eta_*=50$~Mpc, the $\delta T$ profile effectively extends to $\theta\approx 0.015$. The spots are smaller so we increase the resolution to $N_{\rm side} = 512$, leading to a spot radius of $8.5$ pixels and a maximum separation of $6$ pixels. Note that, with these sizes and separations, the profiles of the two hotspots from a given pair will overlap for all benchmark points. Also, as  signals for the $\eta_*=100$ and $50$~Mpc cases are colder and more difficult to search for, we consider an even larger coupling $g=10$ in these two cases when presenting the results.

Some example images from the PHS signal and combination of signal and CMB background for the $\eta_*=160$~Mpc with $g=6$ benchmark are shown below in Fig.~\ref{fig:PHSsignal1}. The images focus on a small patch in the sky within $\approx[-16^{\circ},16^{\circ}]$ for both the longitude and the latitude. The upper left and upper right plots are the PHS+CMB and PHS-only plots made with $N_{\rm side}=256$. To illustrate how $N_{\rm side}$ changes the resolution, the lower left plot is the simulated CMB map, and the lower right plot is the same signal model but made with $N_{\rm side}=2048$. Example images of the $\eta_*=50,\,100$~Mpc scenarios are shown in Appendix~\ref{app.img} (Fig.~\ref{fig:PHSsignal2}~and~\ref{fig:PHSsignal3}).

\begin{figure}[h!]
    \centering
    \includegraphics[width=0.45\textwidth,clip]{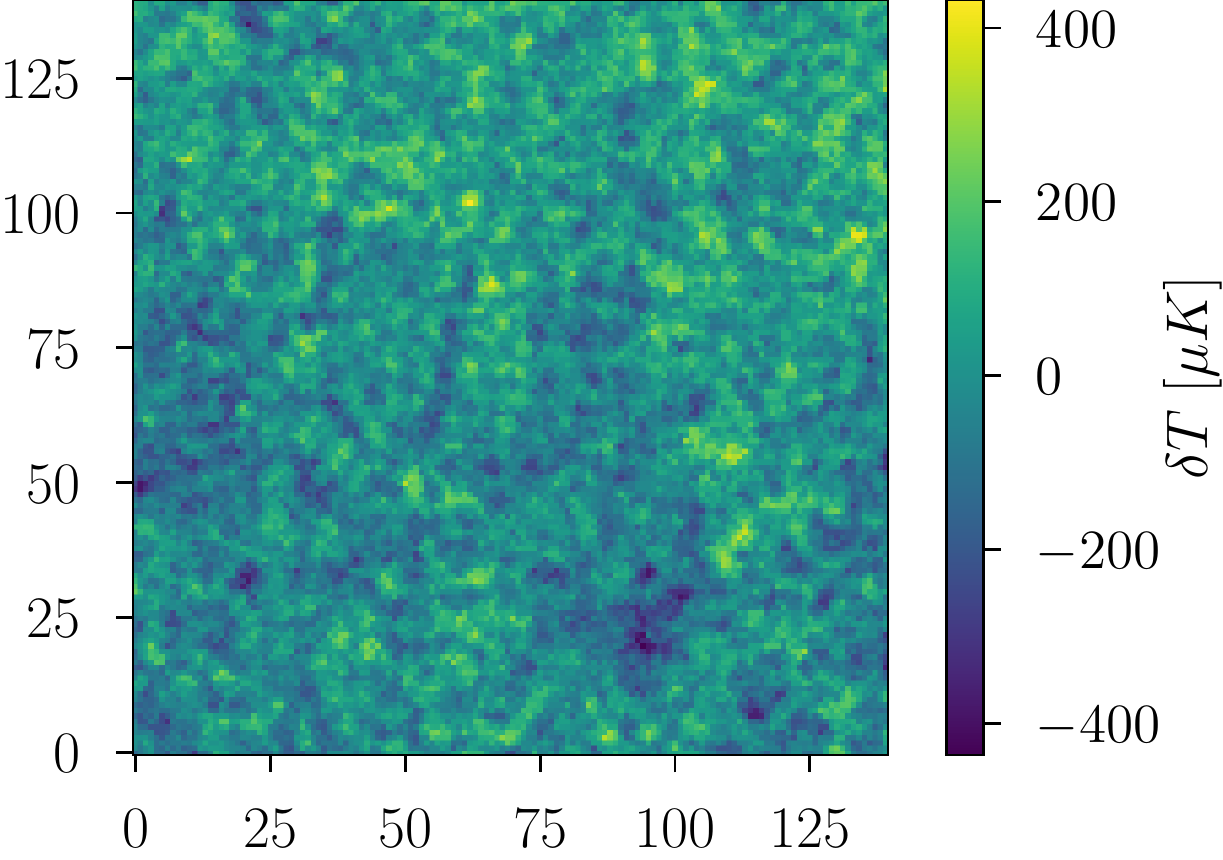}\quad
    \includegraphics[width=0.435\textwidth,clip]{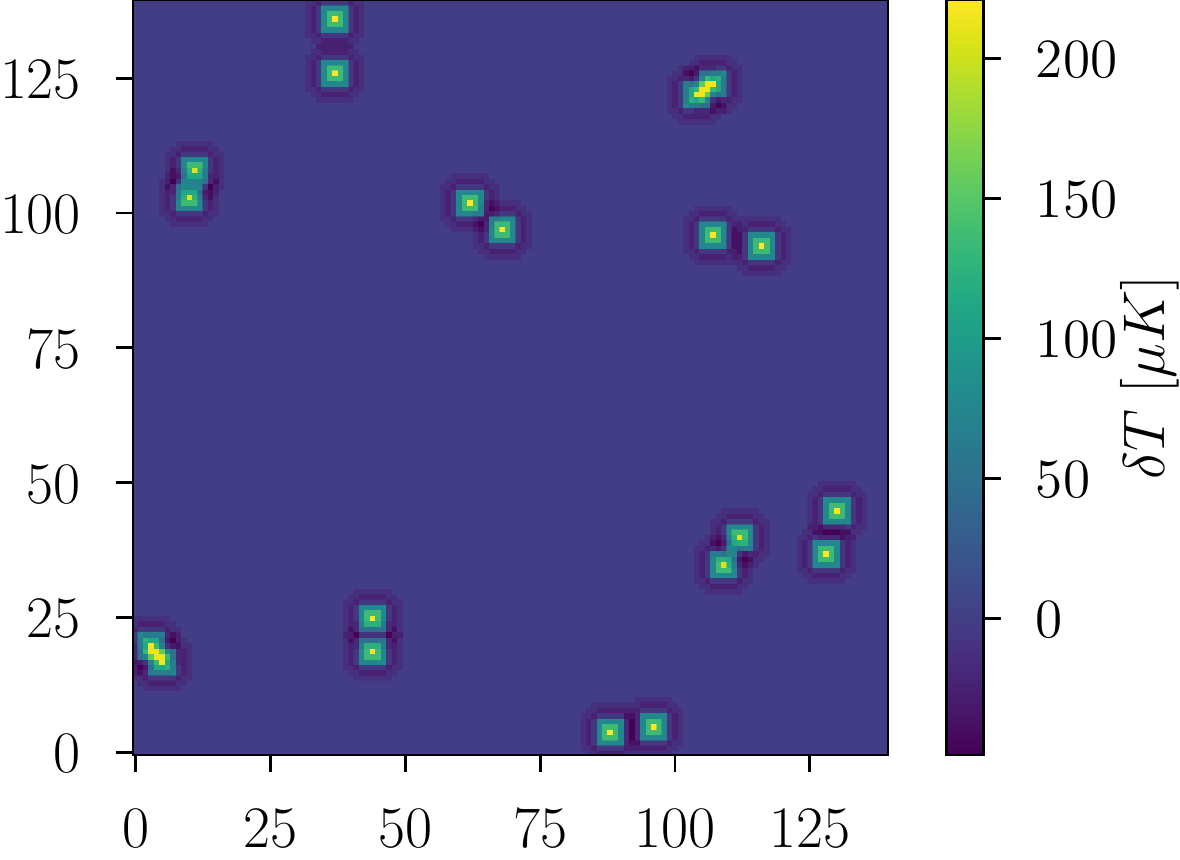}\\
    \includegraphics[width=0.45\textwidth,clip]{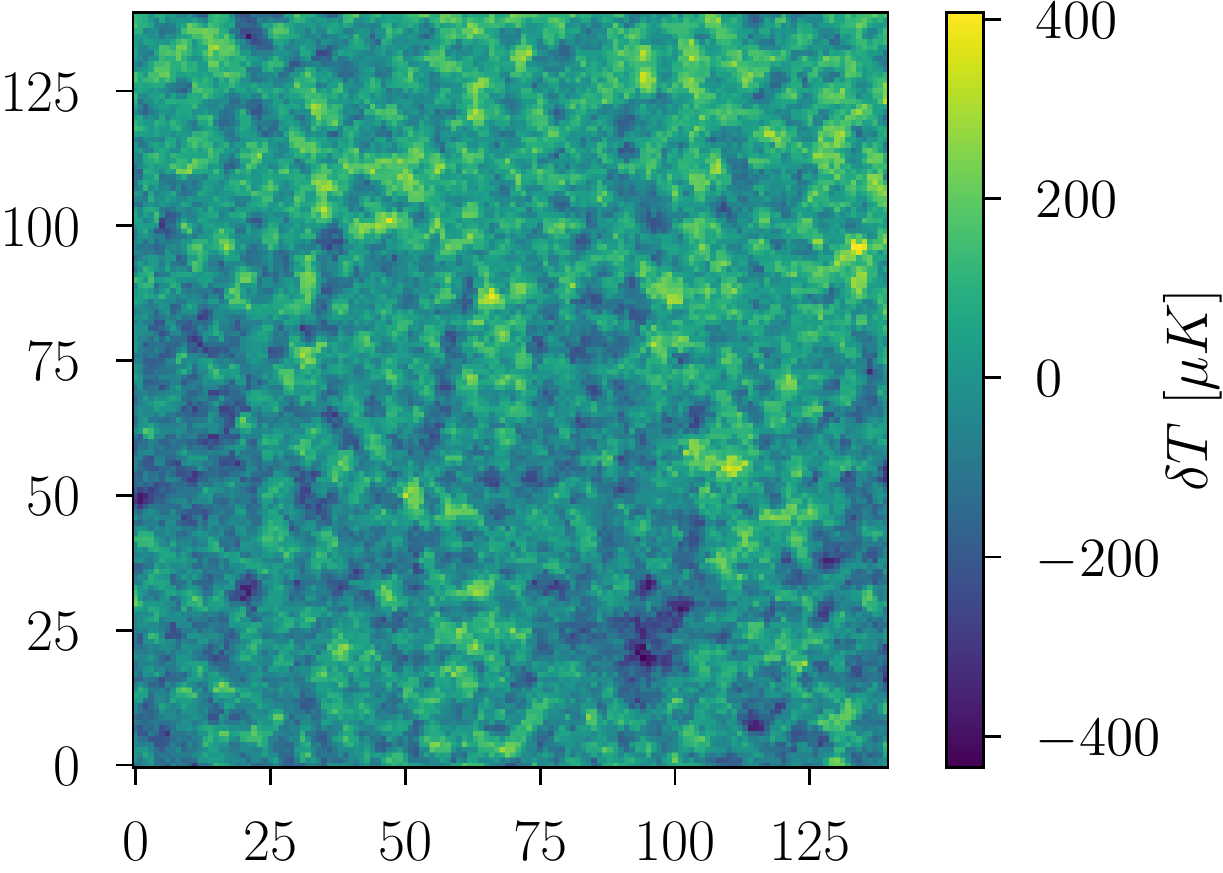}\quad   
    \includegraphics[width=0.435\textwidth,clip]{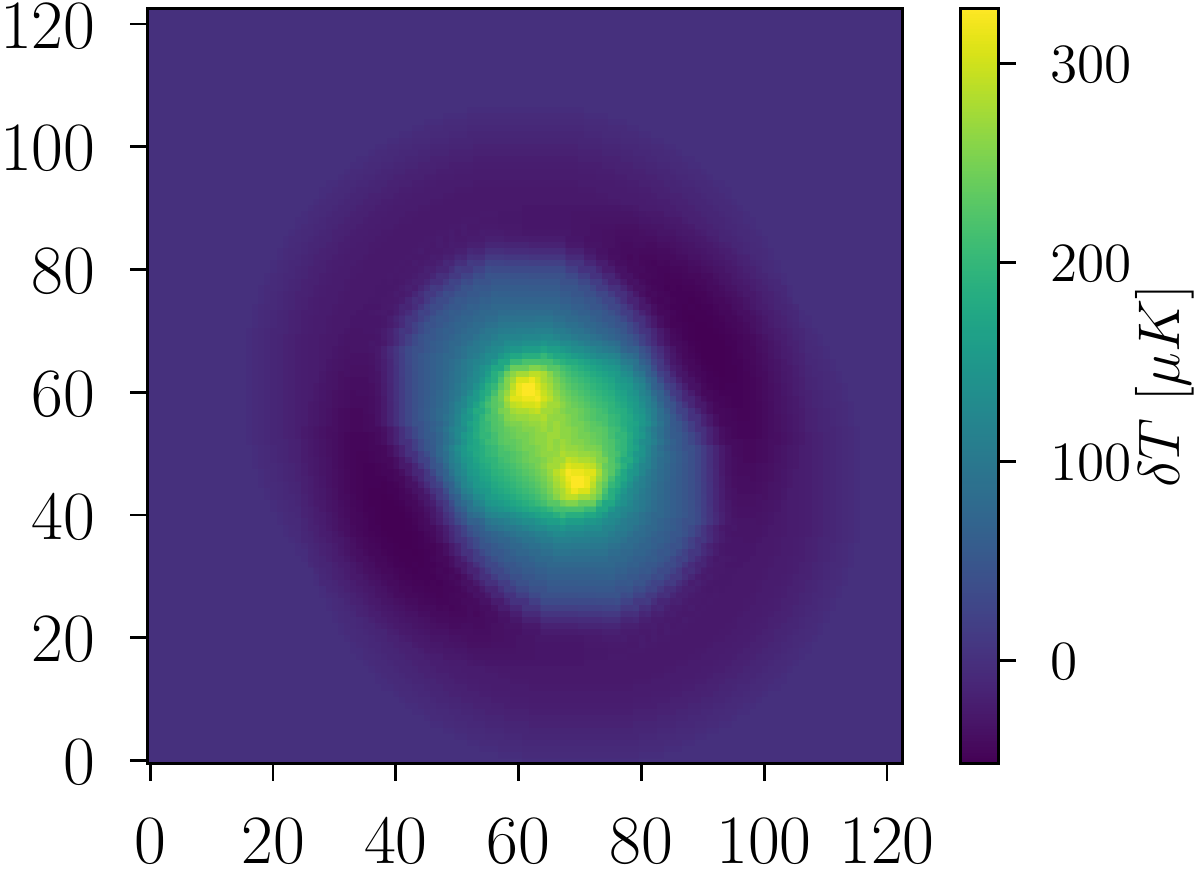}  
    \caption{Local patches of (upper-left) CMB + PHS, (upper-right) PHS, and (lower-left) CMB, generated within a region of longitude = $ [-\ang{16}, \ang{16}$] and latitude = $[-\ang{16}, \ang{16}]$ centered at $(\theta, \phi) = (\pi/2,0)$ and projected onto the Cartesian coordinates ($140 \times 140$ pixels) using $N_{\rm side} = 256$.  The shape of PHS is based on the $\eta_{*} = 160$ Mpc and $g =6$ benchmark, and the separation distance is $r_{\text{sep}} = 0\sim10$ pixels. In total, ten PHS signals are injected using a flat distribution. (lower-right) A higher-resolution image of the PHS with separation $\approx\eta_*/4$ using $N_{\rm side} = 2048$, plotted within a region of longitude = $ [-\ang{2}, \ang{2}$] and latitude = $[-\ang{2}, \ang{2}]$ centered at $(\theta, \phi) = (\pi/2,0)$ projected onto the Cartesian coordinates ($123 \times 123$ pixels).
    }
    \label{fig:PHSsignal1}
\end{figure}
%

\subsection{Searching for PHS signals with temperature cuts}\label{sec.PHSsearch}

From the signal panels in Fig.~\ref{fig:PHSsignal1}, the PHS signal looks very similar to jets at the Large Hadron Collider --  localized regions on a 2D plane with $\sim$ circular shapes that stand out as hotter than the background (more energy/transverse momentum in the case of jets). That analogy motivates us to try a position-space search for the PHS using cuts on temperature, inspired by how jets are hunted for at colliders.

Rather than searching the entire sky all at once for PHS, we divide the CMB maps into patches, then seed them with PHS signals. As part of this process, the patches are projected from spherical coordinate maps to Cartesian coordinates. This procedure -- cutting the sky into `flattened' patches -- is not technically necessary but is often adopted in the literature~\cite{Lopez-Caniego:2007lsc,Lopez-Caniego:2009egp,Ramos:2010bb,Herranz:2008yr,Massardi:2008fi}. Historically it was used since CMB analyzers using wavelets or machine learning could not cope with spherical coordinates~\cite{Gonzalez-Nuevo:2006ood,Lopez-Caniego:2006kci,Sanz:1999xu,Ciuca:2017gca,Montefalcone:2020fkb,Han:2021unz}. This technicality has been overcome in the neutral network analysis~\cite{Perraudin:2018rbt,Fluri:2018hoy}, but we will stick to the flat patches for simplicity in this analysis. Projecting from spherical to Cartesian coordinates does introduce some degree of distortion. To mitigate this effect, we restrict the simulated CMB patches to a region around the equator made by $140\times140$ pixels. For the $\eta_*=160$ Mpc case, this patch is within $\approx[-16^{\circ},16^{\circ}]$ for longitude and latitude and corresponds to roughly $1/40^{\rm th}$ of the sky.  For the $\eta_*=50\,\,(100)$ Mpc  case, this pixel region is within $\approx[-8^{\circ},8^{\circ}] ([-16^{\circ},16^{\circ}])$ longitude and latitude and corresponds to $1/160^{\rm th}\, (1/40^{\rm th})$ of the sky.

Taking a sample $140 \times 140$ pixel patch, we implement the following cuts to isolate the PHS signal from the CMB background:
\begin{itemize}
    \item Reduce $\delta T_{\rm CMB}$ contribution from lower $\ell$-modes by cutting the image into regions with a certain $\ell\lsim\ell_*$, calculate the average temperature of each region and subtract it from the image.
    \item Further veto pixels below a temperature cut $\delta T_{\rm cut}$ that is chosen based on the PHS temperature.
\end{itemize}

\begin{figure}
    \centering
        \includegraphics[width=0.33\textwidth,clip]{Plots/Low_Resol_BKG_Nside256.pdf}
    \includegraphics[width=0.33\textwidth,clip]{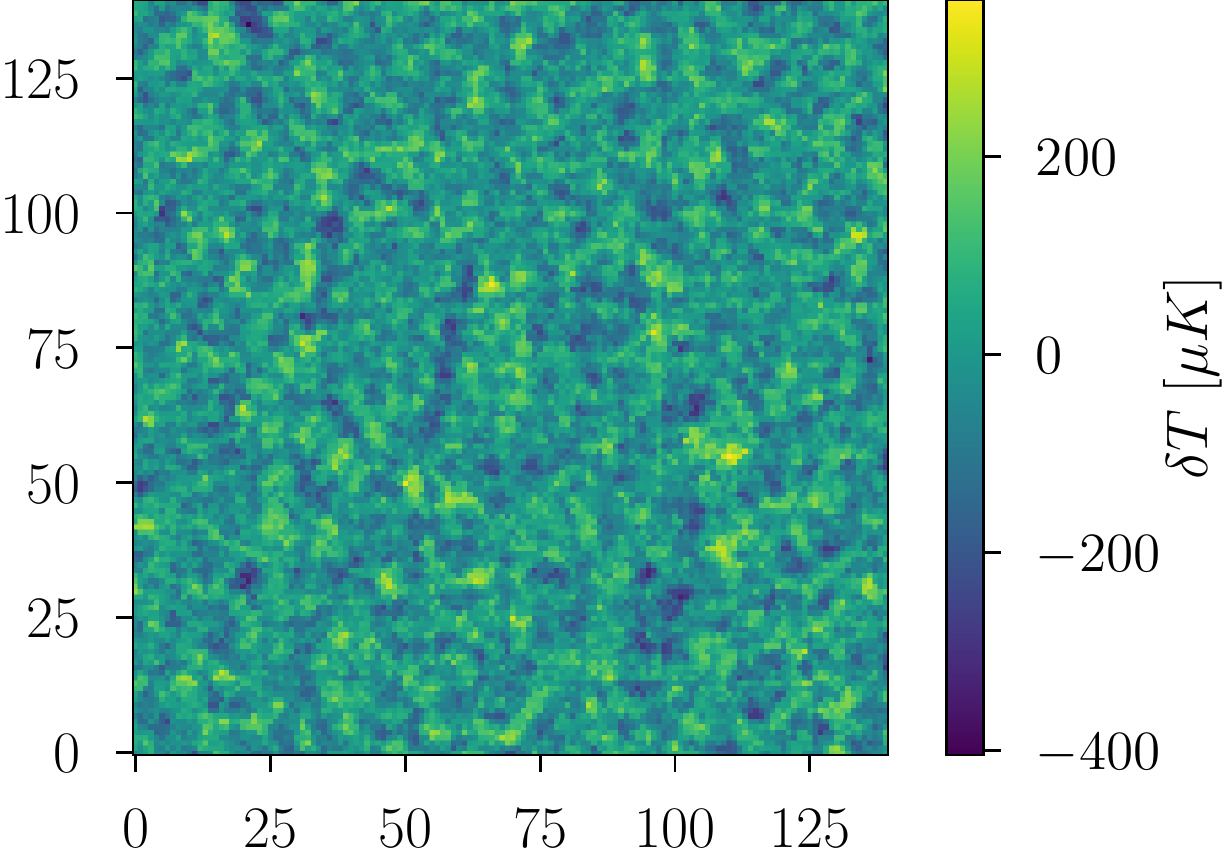}  
    \includegraphics[width=0.32\textwidth,clip]{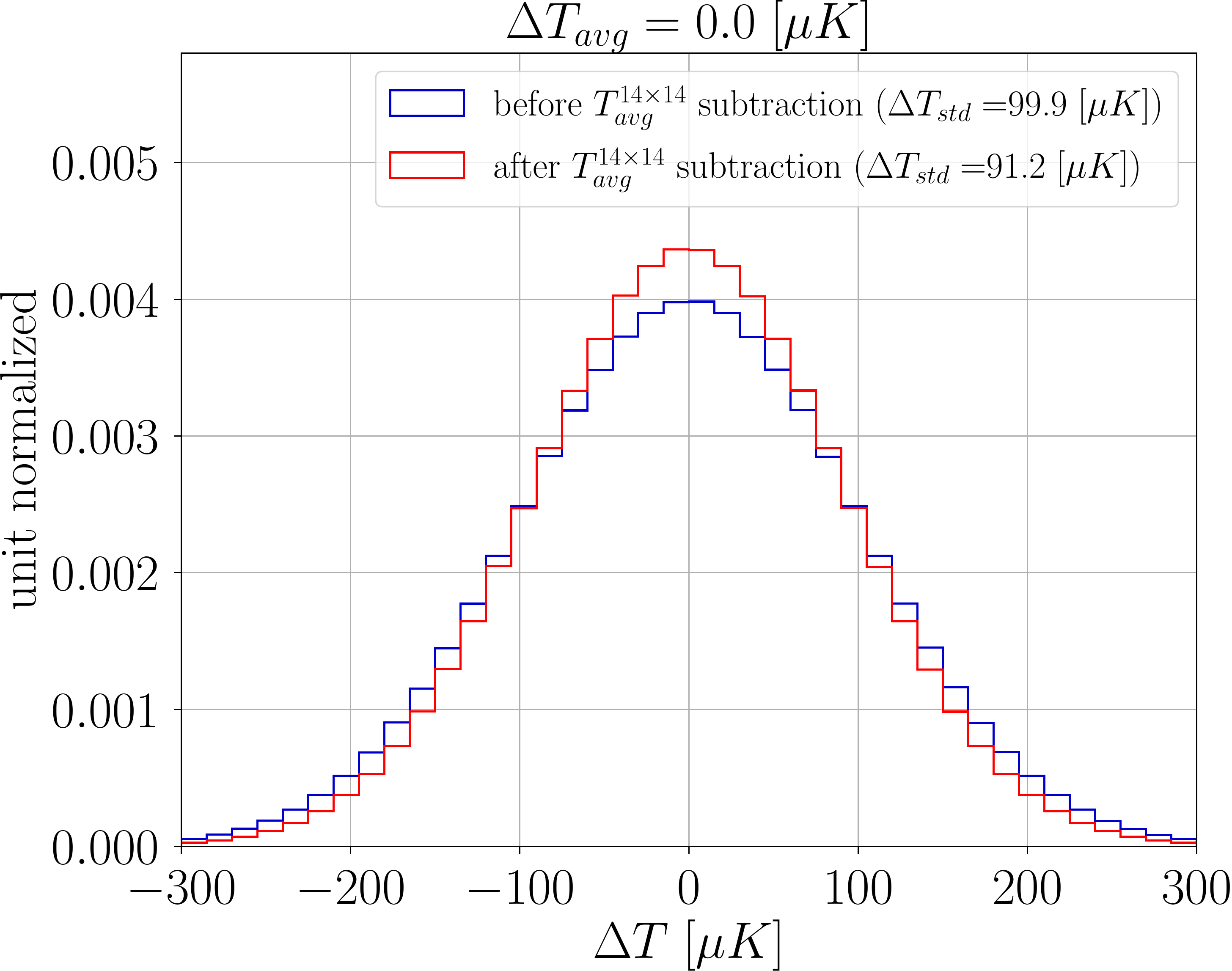}
        \caption{A local patch of (upper-left) CMB within a region of longitude = $ [-\ang{16}, \ang{16}$] and latitude = $[-\ang{16}, \ang{6}]$ centered at $(\theta, \phi) = (\pi/2,0)$ and projected onto the Cartesian coordinates ($140 \times 140$ pixels) using $N_{\rm side} = 256$. (middle) We subtract average temperatures $T^{14\times14}_{avg}$ from smaller regions ($14 \times 14$ pixels) of the patch. (right) The resulting unit-normalized  $\Delta T$ distributions of $10^3$ CMB maps with and without $T^{14\times14}_{avg}$ subtractions.
        }
    \label{fig:lowlcut}
\end{figure}

Even though the temperature of PHS we consider in Eq.~\eqref{eq.sigl} is typically much larger than the primordial $\delta T\approx 27\,\mu$K of a {\it single} $\ell$ mode before entering the horizon, the total standard deviation -- formed from summing over a range of $\ell$ -- is much larger. Specifically,
\begin{equation}
    \sigma^2_{\rm CMB}=\langle\delta T_{\rm CMB}\delta T_{\rm CMB}\rangle=\sum_{\ell_{\rm min}}^{\ell_{\rm max}}\frac{2\ell+1}{4\pi}C_{\ell}^{\rm TT},
\end{equation}
where $\ell_{\rm min}$ is set by the size of the patch and $\ell_{\rm max}$ is set by the pixel resolution, roughly $\ell_{\rm max}\approx\sqrt{12N_{\rm side}^2}$. For the range of $\ell$ for our benchmark signals, $\sigma_{\rm CMB} \sim \mathcal O(110\,\mu K)$, comparable to the PHS signal temperatures $\sim \frac{g}{2}\delta T$ with $g=\mathcal{O}(1)$. To reduce background coming from modes with $\ell$ much smaller than the hotspot signals ($\ell_*$), we divide the simulated maps into regions that correspond to modes with $\ell\lsim\ell_*$,  calculate the average temperature of each division and subtract it from the original map. As is shown in the rightmost panel of  Fig.~\ref{fig:lowlcut}, this process reduces the standard deviation of the $\delta T_{\rm CMB}$ distribution and helps to suppress the background. Depending on the benchmark this subtraction process leads to $\mathcal{O}(10-30)\%$ improvement in the final constraints on the allowed number of hotspots. As a specific example, for our $N_{\rm side}=256$ simulations we divide the $140 \times 140$ pixel map into $14 \times 14$ pixel blocks. Subtracting the average temperature in these sub-patches raises $\ell_{\rm min}$ from $\approx 6$ to $\approx 60$, and lowers $\sigma_{\rm CMB}$ to $91\, \mu$K.

After subtracting contributions from the lower-$\ell$ modes, we apply a $\delta T_{\rm cut}$ to image in the $140\times 140$ pixel region.  The $T_{\rm cut}$ is chosen to optimize the signal significance -- to be defined in more detail below -- but ranges from $140\, \mu K$ to $340\, \mu K$ depending on the benchmark point. For such a simple cut, the number of background spots $\langle N_{\rm CMB}\rangle$ in the region can be estimated analytically by a Gaussian distribution of temperature fluctuations. For example, in the $\eta_*=160$~Mpc analysis, the standard deviation of the $\delta T$ fluctuation is $\approx 91\,\mu$K after subtracting the low-$\ell$ modes. Once applying $\delta T_{\rm cut}\approx 160\,\mu$K to the simulated image, a Gaussian temperature fluctuation only gives $\approx 4\%$ chance for each single pixel to pass the cut. This corresponds to $\langle N_{\rm CMB}\rangle N_{\rm pat}\approx 3\times 10^4$ for the entire sky, where $N_{\rm pat}=40\,(160)$ is the number of $140\times 140$ pixel regions in the whole sky for the $\eta_*=100,\,160$~Mpc ($50$~Mpc) analysis. The number is close to the result derived from the numerical analysis (and shown in Table~\ref{tab:CutFlow4}).

To determine the significance of a given PHS scenario, we determine:
\begin{equation}
 {\rm Sig}\equiv\frac{(\langle N_{n\,{\rm sig+CMB}}\rangle-\langle N_{\rm CMB}\rangle)\,N_{\rm pat}\,f_{\rm sky}}{\sqrt{\langle N_{\rm CMB}\rangle\,N_{\rm pat}\,f_{\rm sky}}}\,,
 \label{eq:signif}
\end{equation}
where $\langle N_{n\,{\rm sig+CMB}}\rangle$ and $\langle N_{\rm CMB}\rangle$ are the average number of hotspots identified in the $140\times 140$ pixel maps with $n$-signal injections and no injections. $f_{\rm sky}=60\%$ is the sky fraction that we assume for the analysis. We determine $\langle N_{n\,{\rm sig+CMB}}\rangle$ by generating $10^3$ CMB background maps and increasing the number of injected PHS signals  until Eq.~\eqref{eq:signif} equals 1 or 2 (for $1\sigma$ or $2\sigma$, respectively) for a given $T_{\rm cut}$. We then iterate the entire process, varying $T_{\rm cut}$ until we find the optimal value. 

The optimized temperature cuts for each benchmark PHS point are shown in Table~\ref{tab:CutFlow4}. We also display the number of PHS pairs that must be injected for each benchmark in order to achieve $\rm{Sig} = 1$ $(\langle N_{1\sigma} \rangle)$  or $\rm{Sig} = 2$ $(\langle N_{2\sigma}\rangle)$. In other words, $\langle N_{1,2\sigma} \rangle$ is the maximum number of PHS  pairs in the entire sky that can hide (at 1 or 2 $\sigma$) from our search. To get an idea of the signal efficiency under the temperature cut, one can compare $2\langle N_{1,2\sigma} \rangle$ to $\langle N_{n\,{\rm sig+CMB}}\rangle - \langle N_{\rm CMB}\rangle$. Depending on the benchmark point, we find the efficiency (per PHS pair) ranges from $\mathcal O(5\%)$ ($\eta_* = 50~\text{Mpc},\, g = 3$) to $\mathcal O(70\%)$ ($\eta_* = 100~\text{Mpc},\, g = 10$), see Table~\ref{tab:CutFlow4}. From Table~\ref{tab:CutFlow4}, this simple cut and count search is sensitive to  $\mathcal{O}(10-10^4)$ PHS, a number that the scalar model discussed in Sec.~\ref{sec.back} can easily generate.

One signal feature that is obviously missing from our analysis is the pairwise nature of the hotspots. Viewing the signal on its own, as in Fig.~\ref{fig:PHSsignal1}, the fact that there are always two spots in close proximity is striking, and it seems like one should be able to exploit this to increase the power of a position-space search. Taking a cue from jet physics once again, one simple way to capture this pairwise feature is impose an additional cut requiring any pixel hotter than $\delta T_{\rm cut}$ to have a second hot pixel (either at the same temperature or slightly different) within some neighborhood in order to be considered a viable signal candidate. In practice, we find that a search looking for \textit{pairs} of hot pixels within disc-like regions on the sky fares worse than the simple $\delta T_{\rm cut}$ search, even when optimized over the distance between hot pixels and the individual pixel temperature cuts. Part of the difficulty with the paired pixel search is that the core temperature of the PHS -- found by multiplying the profile in Fig.~\ref{fig:hsprofile} by $g$ -- is not significantly bigger than $\sigma_{\rm CMB}$. For example, for $g = 3$, the core temperature of the $\eta_* = 160$ Mpc hotspots is $\mathcal O(120\, \mu K)$, while for the $\eta_* = 50, 100$ Mpc benchmarks the core temperature is only $\mathcal O(30\, \mu K)$. With PHS temperatures so close to $\sigma_{\rm CMB}$, it is impossible to find $\delta T_{\rm cut}$ values for two pixels that are both efficient for the signal and not plagued by high background fake rate. Dropping the requirement for two hot pixels helps because we can raise the single pixel $\delta T_{\rm cut}$; while the individual PHS may not be particularly hot compared to $\sigma_{\rm CMB}$, the profiles overlap, so the odds of one pixel passing an increased $\delta T_{\rm cut}$ remain high while the background rate falls precipitously. A more sophisticated analysis -- perhaps using template fitting or a machine learning to better utilize the shape of the profiles -- could enhance the signal sensitivity significantly, and we will pursue this in a future study.

\begin{table}[h]
\begin{center}
\renewcommand{\arraystretch}{1.2}
\scalebox{1}{
\begin{tabular}{|c|c|c|c|c|}
\hline
     $\eta_* = 160$~Mpc         &       $g=2$            &       $g=3$            &       $g=6$         
     \\  
\hline \hline
$\delta T_{\rm cut}$          &    $160 \mu K $    &    $200 \mu K $    &  $340 \mu K $   \\  
$\langle N_{\rm CMB}\rangle N_{\rm pat}$  &   31160                &   11121                &  78                       \\ 
$\langle N_{1\sigma} \rangle$  &   313                &   139                & 16                      \\   
$\langle N_{2\sigma} \rangle$  &   635                &   274                &  48                 \\   
\hline
\end{tabular}} \\[.2cm]

\scalebox{1}{
\begin{tabular}{|c|c|c|c|c|c|}
\hline
     $\eta_* = 100$~Mpc      &       $g=3$         &       $g=6$      &         $g=10$       \\  
\hline \hline
$\delta T_{\rm cut}$           &     $140 \mu K $    &    $160 \mu K $  &    $200 \mu K $     \\   
$\langle N_{\rm CMB}\rangle N_{\rm pat}$   &   48980                 &   31160              &   11121              \\
$\langle N_{1\sigma} \rangle$                    &  1300                   &    301             &  99           \\   
$\langle N_{2\sigma} \rangle$                    &  2681                   &    575             &  207           \\   
\hline
\end{tabular}} \\[.2cm]
\scalebox{1}{
\begin{tabular}{|c|c|c|c|c|c|}
\hline
     $\eta_* = 50$~Mpc        &       $g=3$         &       $g=6$      &         $g=10$          \\  
\hline \hline
$\delta T_{\rm cut}$            &      $140 \mu K $   &   $160 \mu K $   &    $180 \mu K $         \\   
$\langle N_{\rm CMB}\rangle N_{\rm pat}$    &   183730                 &   115295              &   69321                     \\
$\langle N_{1\sigma} \rangle$                     &  7045                   &  1823           &  653       \\   
$\langle N_{2\sigma} \rangle$                     &  14656                   &  3920           &  1335       \\   
\hline
\end{tabular}} 
\caption{Optimized $\delta T_{\rm cut}$ values for each benchmark scenario, along with the number of background events $\langle N_{\rm CMB}\rangle N_{\rm pat}$ and the number of injected hotspot pairs to a achieve $1$ or $2 \sigma$ excess following the Eq.~\eqref{eq:signif} ($\langle N_{1\sigma} \rangle, \langle N_{2\sigma}\rangle $ respectively).}
\label{tab:CutFlow4}
\end{center}
\end{table}

\begin{table}[h]
\begin{center}
\renewcommand{\arraystretch}{1.2}
\scalebox{1}{
\begin{tabular}{|c|c|c|c|c|}
\hline
     $\eta_* = 160$~Mpc         &       $g=2$            &       $g=3$            &       $g=6$         
     \\  
\hline \hline
$\langle N_{\Delta \chi^2 = 6} \rangle$  &   850                &   430               & 110                      \\   
$\langle N_{\Delta \chi^2 = 20} \rangle$  &   1740                &   760                & 190                 \\   
\hline
\end{tabular}} \\[.2cm]

\scalebox{1}{
\begin{tabular}{|c|c|c|c|c|c|}
\hline
     $\eta_* = 100$~Mpc      &       $g=3$         &       $g=6$      &         $g=10$       \\  
\hline \hline
$\langle N_{\Delta \chi^2 = 6} \rangle$                    &  1430                   &    380             &  150           \\   
$\langle N_{\Delta \chi^2 = 20} \rangle$                    &  2640                   &    680             &  240           \\   
\hline
\end{tabular}}
\quad\scalebox{1}{
\begin{tabular}{|c|c|c|c|c|c|}
\hline
     $\eta_* = 50$~Mpc        &       $g=3$         &       $g=6$      &         $g=10$          \\  
\hline \hline
$\langle N_{\Delta \chi^2 = 6} \rangle$                     &  2730                   &  690           &  250       \\   
$\langle N_{\Delta \chi^2 = 20} \rangle$                     &  4940                   &  1270           &  460       \\   
\hline
\end{tabular}} 
\caption{
Minimum numbers of signals $\langle N_{\Delta \chi^2 = 6} \rangle$ ($\langle N_{\Delta \chi^2 = 20} \rangle$) in the sky for a $\Delta \chi^2 = 6$ ($\Delta \chi^2 = 20$) excess based on the goodness-of-fit test using $C_{\ell}^{\rm TT}$ distributions.
}
\label{tab:CutFlow5}
\end{center}
\end{table}

\subsection{PHS signatures in the CMB power spectrum and bispectrum}\label{sec.ClTT}

\begin{figure}
\center{
\includegraphics[width=7.0cm]{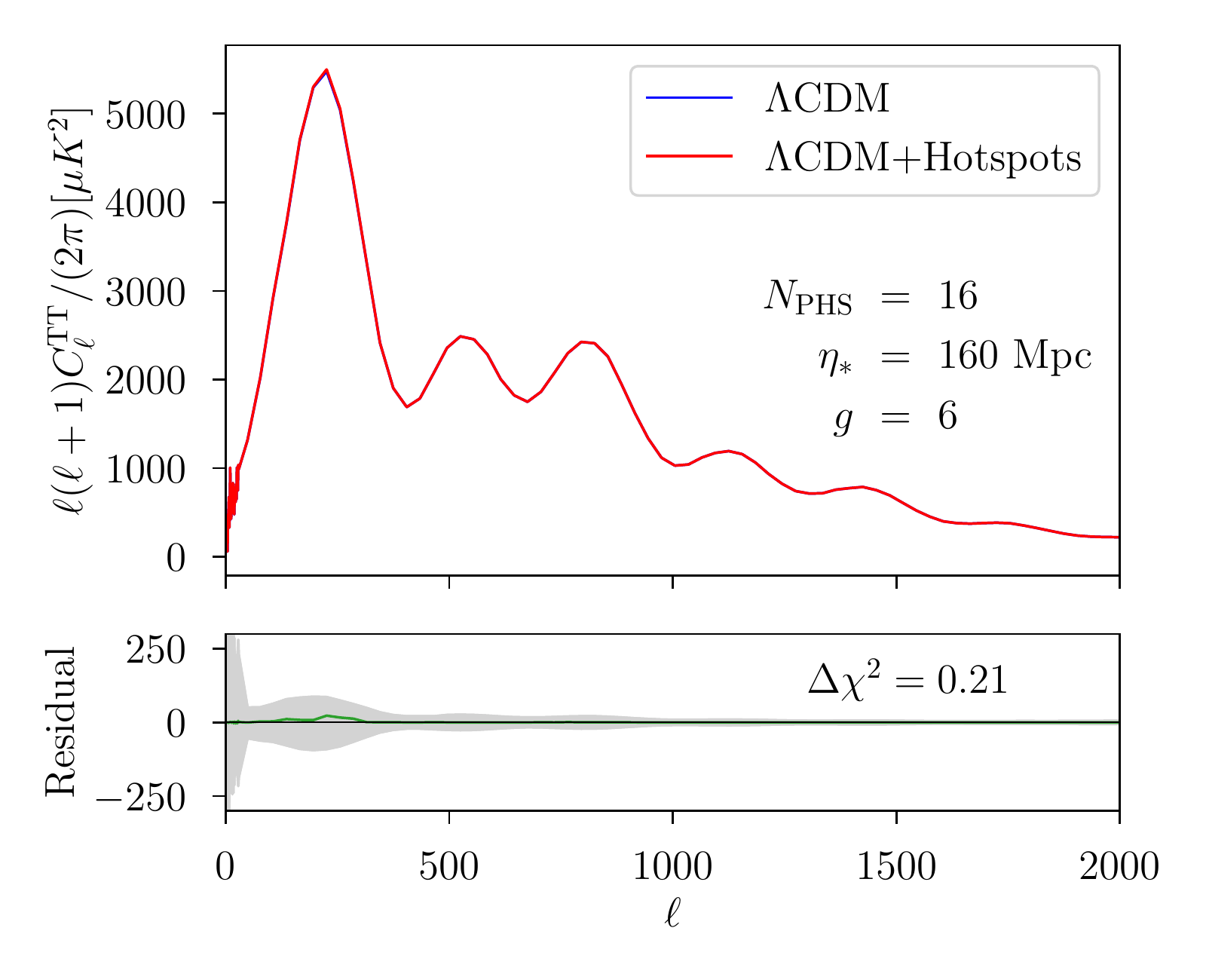}\\
\includegraphics[width=7.0cm]{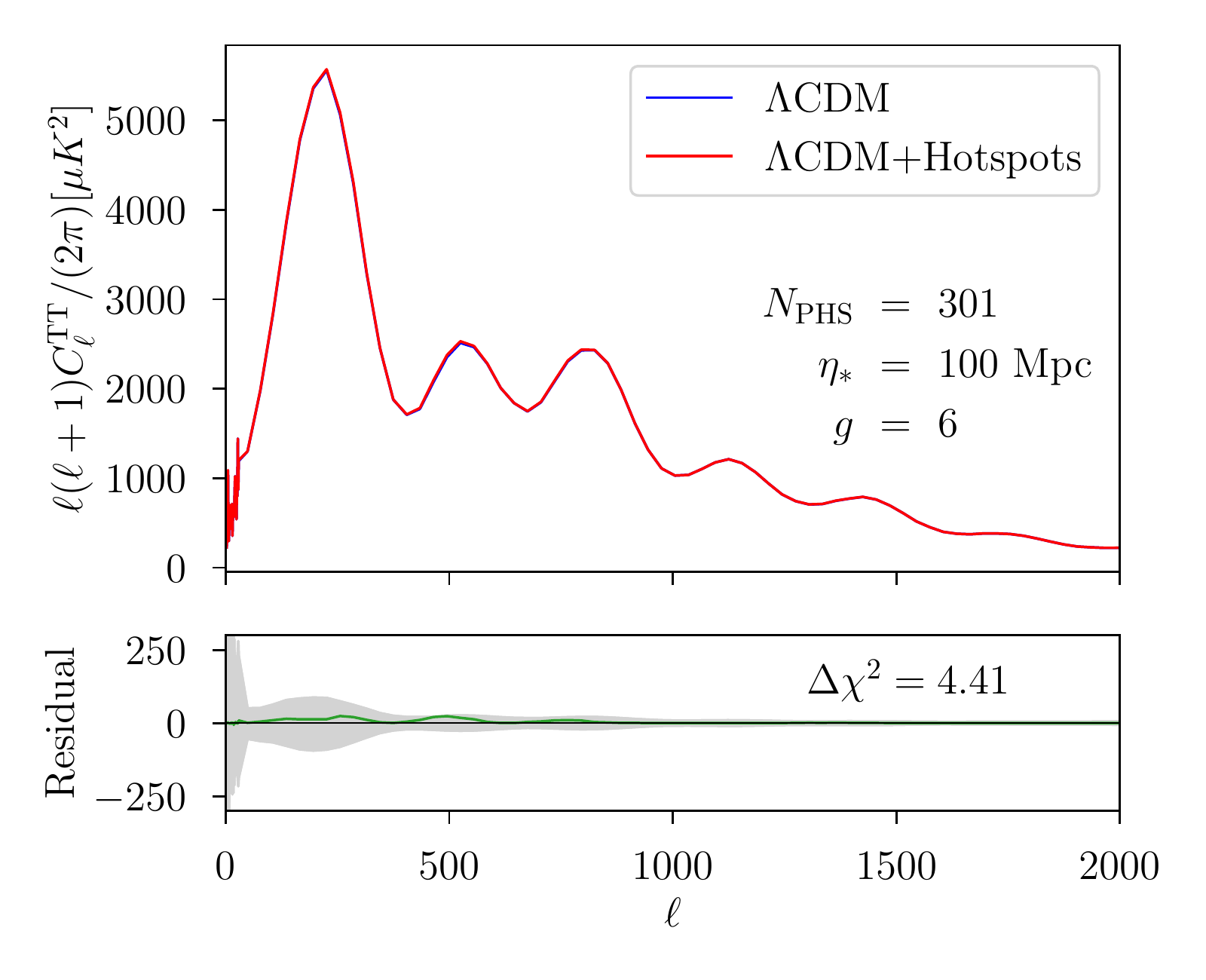}
\includegraphics[width=7.0cm]{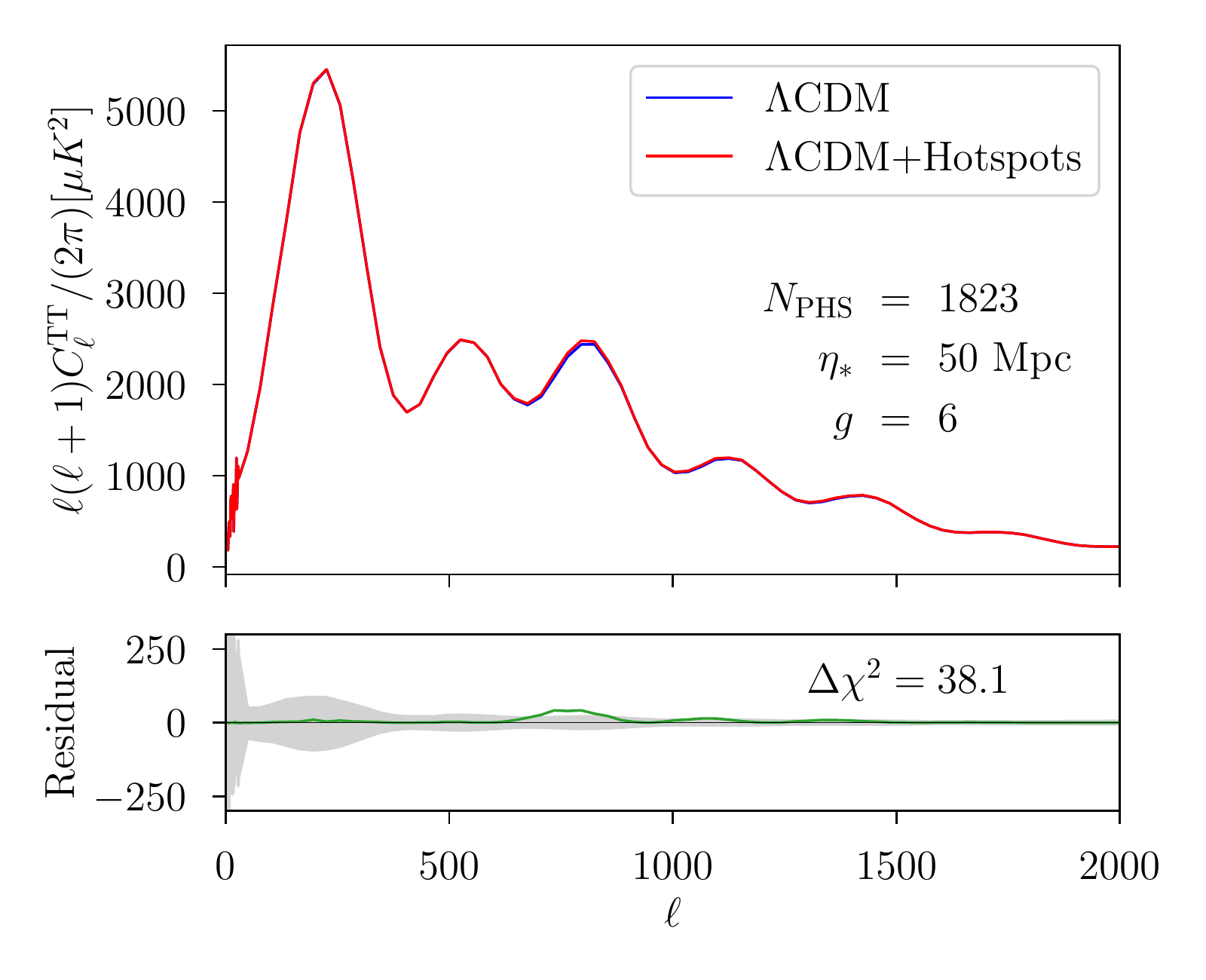}
\caption{
$\mathcal{D}_{\ell}^{\rm TT}$ spectrum of the CMB using best fit $\Lambda \rm CDM$ input parameters in Eq.~(\ref{eq.default}) with (red lines) and without (blue lines) PHS signals injected in the sky using a resolution parameter $N_{\rm side} = 2048$. The differences between the two distributions are shown in green lines, and the shaded regions correspond to $1\sigma$ uncertainty, taken from the Planck 2018 data \cite{Aghanim:2018eyx}. Here we show examples with $g=6$. 
Several benchmarks of $\eta_{*}$  and the number of PHS signals corresponding to $\langle N_{1\sigma} \rangle$ bounds in Table~\ref{tab:CutFlow4} are chosen. The corresponding $\Delta \chi^2$ excess based on the goodness-of-fit test is shown in each panel.
}\label{fig:DlTT}
}
\end{figure}

As we discussed in Sec.~\ref{sec:ClTT}, the existence of PHS gives an overall increase of the $C_{\ell}^{\rm TT}$ spectrum for modes $\ell\lsim \ell_*\approx\eta_0/\eta_*$, and the increase is proportional to the number of hotspots $N_{\rm HS}$ and the heavy scalar-$\phi$ coupling $g^2$. To verify this behavior, we add $N_{\rm HS}$ hotspots to the \texttt{HEALPix} CMB map using the profile from Eq.~\eqref{eq.thprofile} (translated to pixels), then derive the power spectrum. In order to be sensitive to $\ell =2000$, we work with a higher resolution, $N_{\rm side}=2048$. Some example higher resolution images are shown in Figs.~\ref{fig:PHSsignal1}, \ref{fig:PHSsignal2}, \ref{fig:PHSsignal3} (bottom right panels).

In Fig.~\ref{fig:DlTT}, we show the $\mathcal{D}_{\ell}^{\rm TT}$ spectra obtained from the \texttt{HEALPix} for the $\Lambda$CDM model (blue) with the best-fit values of the parameters from Planck~2018 results~\cite{Aghanim:2018eyx}, and also from the $\Lambda$CDM+PHS (red) with the number of PHS that gives a $1\sigma$ excess in the cut-and-count analysis (Table~\ref{tab:CutFlow4}). In the lower panel of each plot, we show the difference between the two scenarios (green) as well as the $1\sigma$ error bars from the Planck~2018 result~\cite{Aghanim:2018eyx} (gray), following the same binning of $\ell$. We also give the total $\Delta\chi^2$ from the green curves in each plots. Our reasoning for showing the deviation from a $1\sigma$ excess (instead of $2\sigma$) of the cut-and-count search is that we would like to compare $\Delta\chi^2$ value to the numbers from the Planck fit which roughly correspond to $1\sigma$ variations of the $\Lambda$CDM parameters. 

Studying Fig.~\ref{fig:DlTT}, we see the green curves roughly follow the behavior expected from Sec.~\ref{sec:ClTT}, though with more complicated $\ell$ dependence coming from the detailed shape of the profile. For $g=6$, we see $<1\sigma$ deviations for all the $\ell$ bins with the $\eta_*=160$ and $100$~Mpc benchmarks. These deviations of the $C_{\ell}^{\rm TT}$ spectra contribute to total $\Delta\chi^2\approx 1-5$. As a comparison, the $\Delta\chi^2$ between the fits of Planck~2018 TTTEEE+lowl+lowE(+lensing) data with the best-fit $\Lambda$CDM parameters and the average parameters is $\approx26\,(6)$ and is $\approx20$ if including only TT+lowl+lowE data~\cite{Planck}. Given that the best fit parameters in these data sets are within the $1\sigma$ variation around the average value of each parameter, the $\Delta\chi^2$ of the deviation in $\mathcal{D}_{\ell}^{\rm TT}$ from PHS signals with a $1\sigma $ excess in the cut-and-count analysis is smaller compared to the $\Delta\chi^2$ between $\Lambda$CDM best fits and average values of Planck data. Therefore the cut-and-count analysis would be able to place a stronger constraint compared to standard TT power spectrum search.  

\begin{figure}
\center{
\includegraphics[width=7cm]{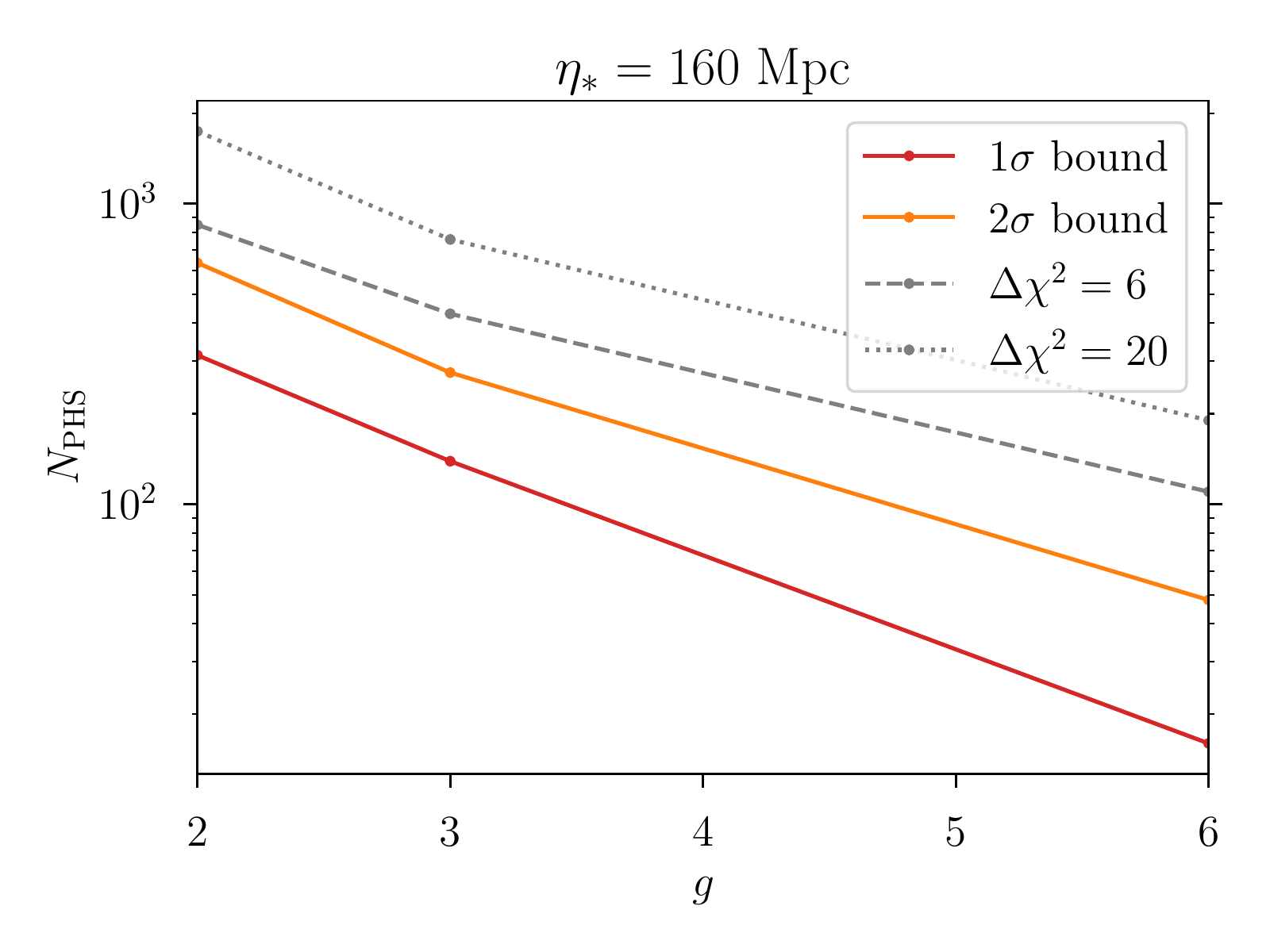}
\includegraphics[width=7cm]{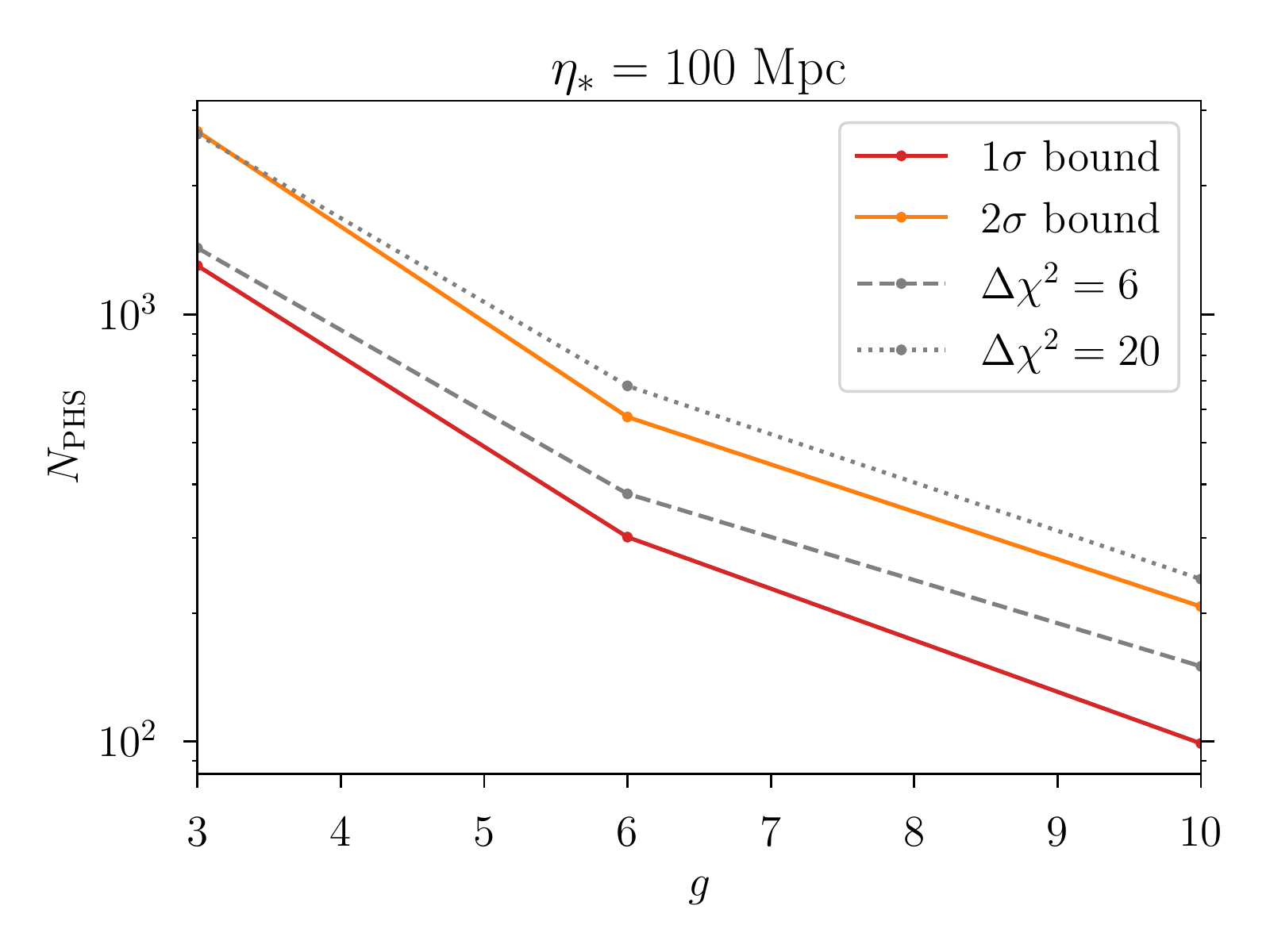}
\includegraphics[width=7cm]{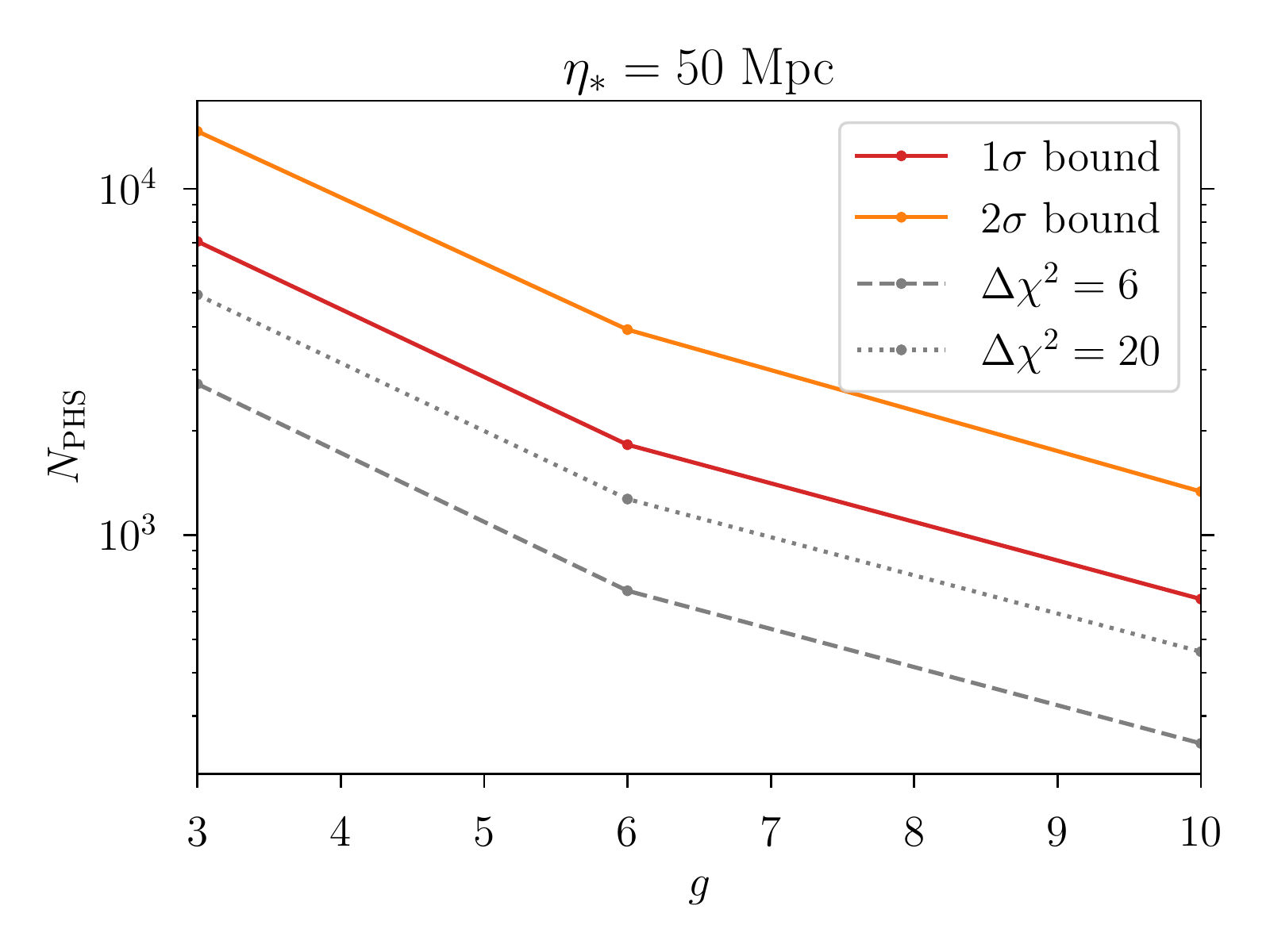}
\caption{Upper bounds on the total number of PHS, $N_{\rm PHS}$ in the entire sky with respect to the heavy scalar-inflaton coupling $g$ for (upper-left) $\eta_* = 160$ Mpc, (upper-right) $\eta_* = 100$ Mpc, and (lower) $\eta_* = 50$ Mpc. We interpolate the results for different benchmarks for the coupling $g$. The $1\sigma$ and $2\sigma$ ($\Delta \chi^2 = 6, 20$) bounds are obtained using the results in Table \ref{tab:CutFlow4} (Table \ref{tab:CutFlow5}). This shows that both for $\eta_*$=160 Mpc and $\eta_*$=100 Mpc, the simple cut-and-count search can perform better than a power spectrum search.}\label{fig:spot_limits}}
\end{figure}

However for $\eta_*=50$~Mpc benchmark, the corresponding $\Delta\chi^2=38$, and the same number of PHS from the cut-and-count study may have been excluded by the power spectrum measurement. To be conclusive, a more dedicated study is necessary to constrain the PHS signal using the Markov Chain Monte Carlo (MCMC) method that would fit the observed data using both the $\Lambda$CDM and PHS parameters. In the present work we do not carry out such an MCMC study, but to get a feeling of the relative sensitivities between the cut-and-count search and the $C_{\ell}^{\rm TT}$ measurement, we calculate the number of PHS signals that corresponds to either $\Delta\chi^2=6$ or $\Delta\chi^2=20$ in each model and show the results in Table~\ref{tab:CutFlow5}. The required $\langle N_{1\sigma}\rangle$ for the cut-and-count search with $g=2\sim6$ are all lower than $\langle N_{\Delta \chi^2 = 6} \rangle$ for the deviation of the TT spectrum. This means the simple cut-and-count study can already provide better constraints to the power spectrum measurement for $\eta_*$=160 and 100~Mpc. A search for bump features with a specific profile using Planck~2015 data and forecasts for LSS surveys appear in \cite{Chen:2016vvw}. It would be very interesting to carry out a similar procedure for the present scenario. The above results from our search for hotspots is summarized in Fig.~\ref{fig:spot_limits}.

In addition to correction to the power spectrum, the presence of PHS also generates primordial non-Gaussianity encoded in the bispectrum and higher point correlation functions, for $k_*\sim1/\eta_*$. From a naive dimensional analysis, the going rate of the magnitude of the dimensionless bispectrum, for a given number density $n$ of PHS per Hubble volume and coupling $g$, is of order $f_{\rm NL}(k_*)\sim 10^{3}g^3(n/H_*^3)$~\cite{Flauger:2016idt}. For $g=2$ and $\eta_*=160$~Mpc, we have $(n/H_*^3)\approx 0.05$, corresponding to the $1\sigma$ result in Table.~\ref{tab:CutFlow4}. This gives $f_{\rm NL}(k_*)\sim 400$. Such a large $f_{\rm NL}(k_*)$, however, only shows up in a narrow range of $k$-modes around $k_*\sim1/\eta_*$. As was found in~\cite{Flauger:2016idt}, for similar models of particle production, the overlap between a large three point functions with a narrow $k$-range and the standard equilateral template but with a wider $k$-range can be small, and new templates are necessary to pick up the non-Gaussian signal. A careful study involving $N$-point correlation functions (allowing for $N>4$) of the model in~\cite{Flauger:2016idt} was performed in~\cite{Munchmeyer:2019wlh} using the WMAP results. To make a rough comparison, we can consider small values of $\omega$ (frequency of particle production) in Fig.~7 of~\cite{Munchmeyer:2019wlh}, as appropriate for our scenario of particle production at a single instant and use the relation,
\begin{align}
\frac{n}{H_*^3} = \frac{1}{2\pi}\left(\frac{\eta_*}{\eta_0}\right)^3\frac{\eta_0}{\Delta\eta_{\rm rec}}N_{\rm PHS},   
\end{align}
to relate the quantity $n/H_*^3$ used by~\cite{Munchmeyer:2019wlh} and $N_{\rm PHS}$ in  Fig.~\ref{fig:spot_limits}.
Using the upper bounds on $N_{\rm PHS}$ in Fig.~\ref{fig:spot_limits}, we then see the constraints derived using our simplified search is better than the sensitivity quoted in~\cite{Munchmeyer:2019wlh}.
We need to improve both the positions space search and the apply $N$-point function searches to the present context to better understand the complementarity between the two types of studies.

\subsection{Bounds on the heavy scalar production}

\begin{figure}
\center{
\includegraphics[width=7cm]{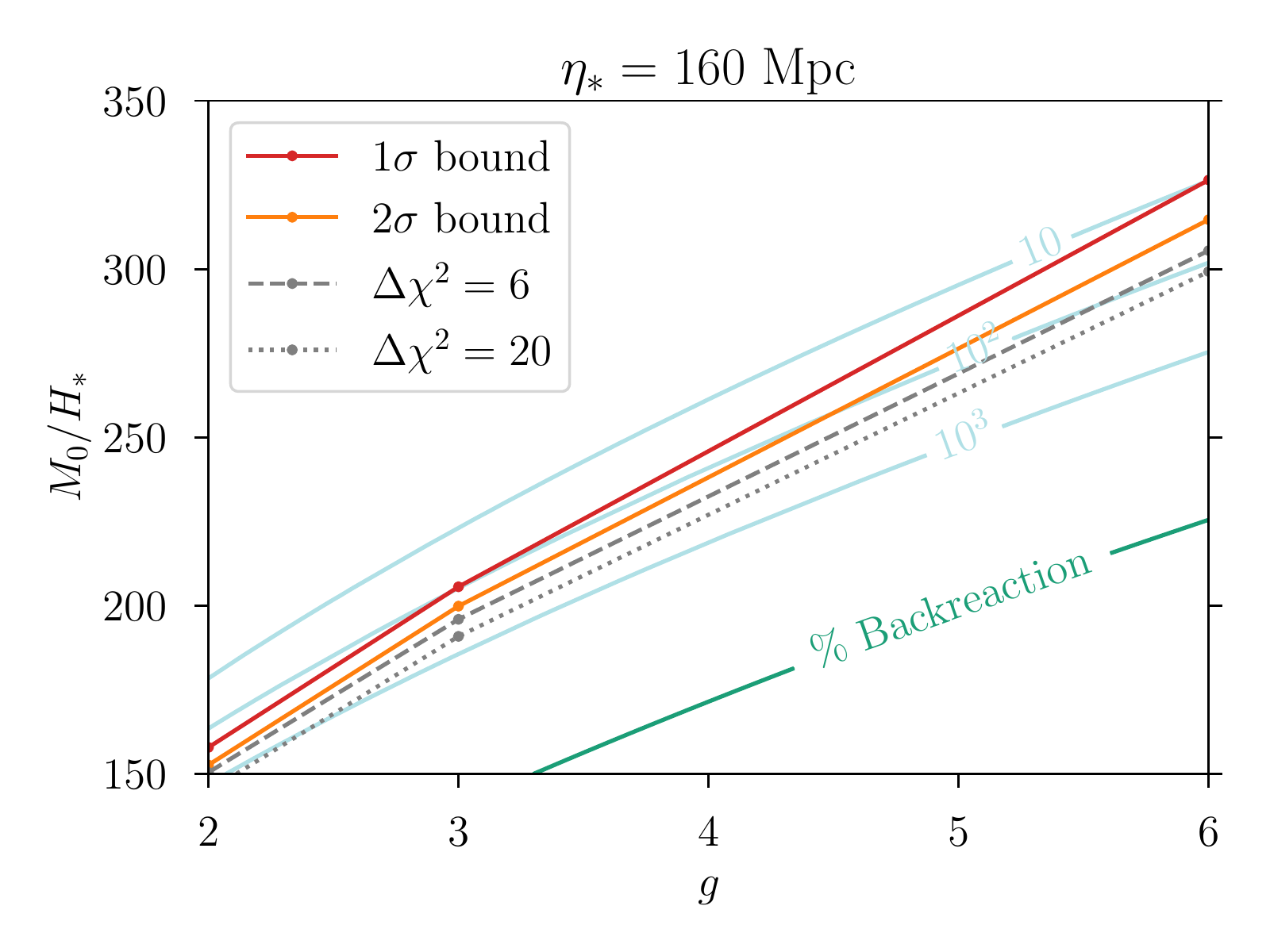}
\includegraphics[width=7cm]{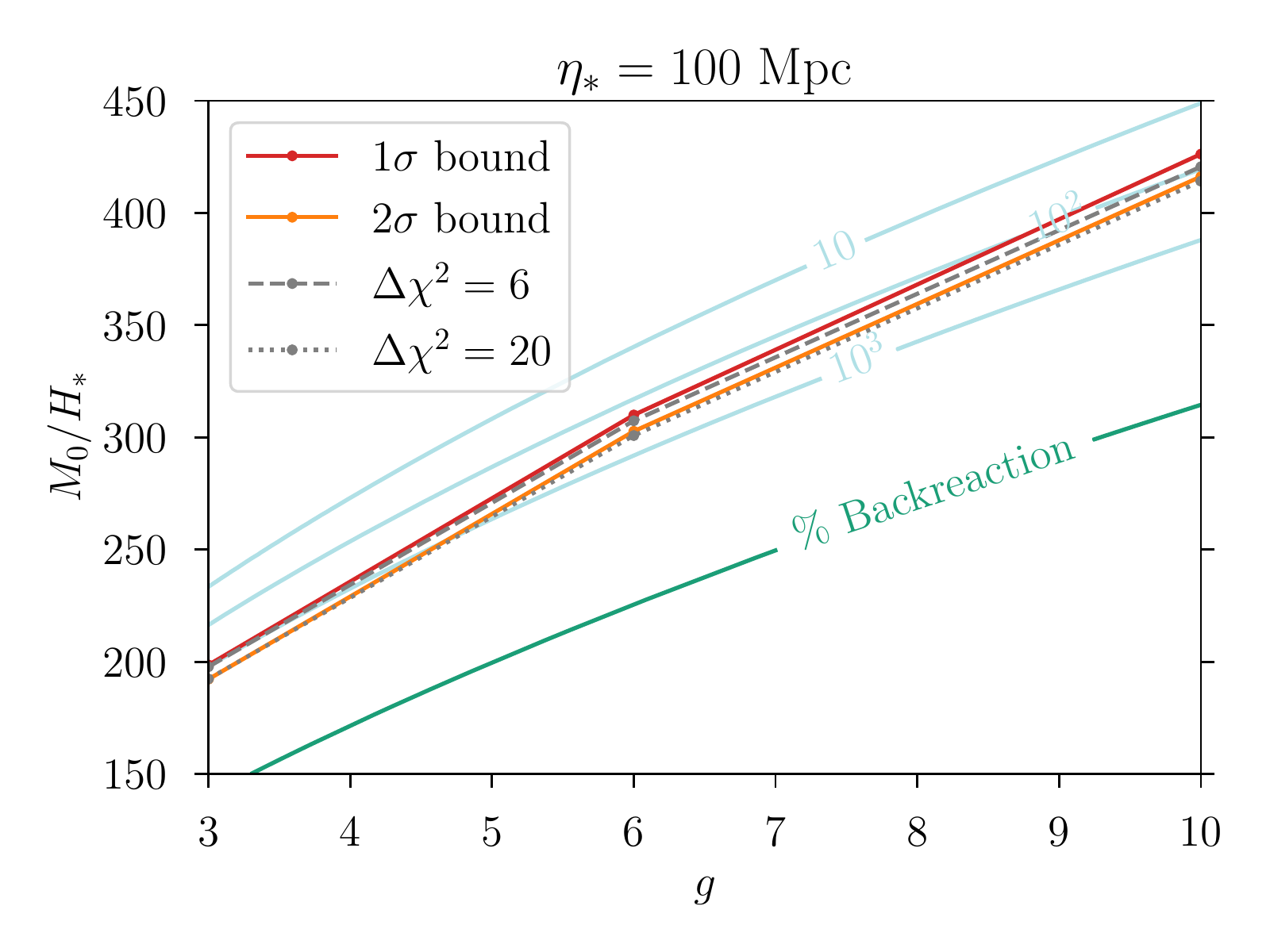}
\includegraphics[width=7cm]{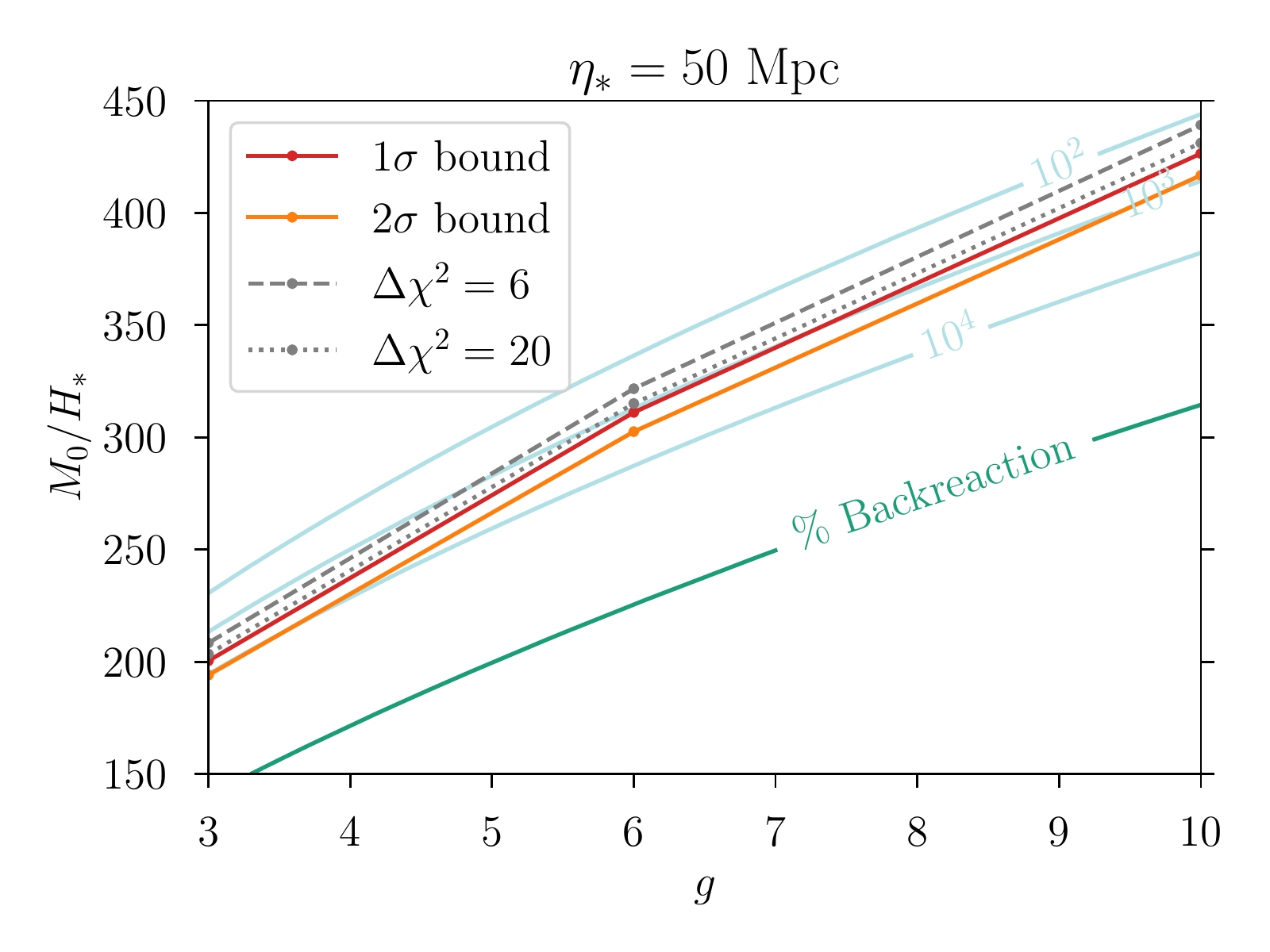}
\includegraphics[width=7cm]{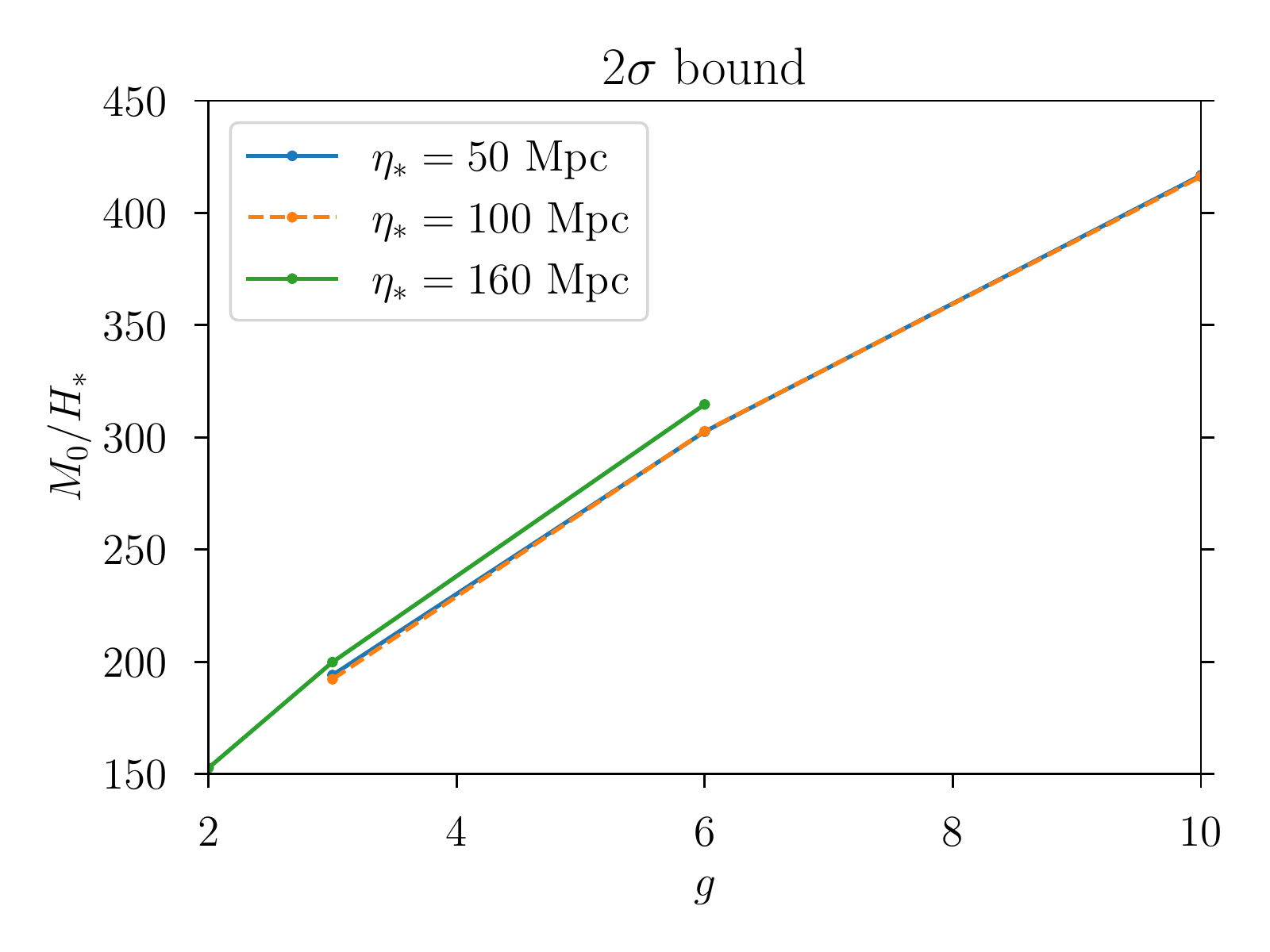}
\caption{Lower bounds on $M_0/H_*$ with respect to the heavy scalar-inflaton coupling $g$ for (upper-left) $\eta_* = 160$ Mpc, (upper-right) $\eta_* = 100$ Mpc, and (lower-left) $\eta_* = 50$ Mpc. We interpolate the results for different benchmarks for the coupling $g$. The $1\sigma$ and $2\sigma$ ($\Delta \chi^2 = 6, 20$) bounds are obtained using the results in Table \ref{tab:CutFlow4} (Table \ref{tab:CutFlow5}). We compare the $2\sigma$ bounds for different production times $\eta_* = 160,100, 50$ Mpc in the lower-right panel. The results for $\eta_*=50$~Mpc and $\eta_*=100$~Mpc cases are almost identical and the corresponding bounds overlap.}\label{fig:limits}}
\end{figure} 

Using Eq.~(\ref{eq.nhotspotsfinal}), we can translate the $\langle N_{1,2\sigma}\rangle$ values in Table~\ref{tab:CutFlow4} to the minimum mass of heavy scalar $M_0$ at the time of particle production. We show the results as the red ($1\sigma$) and orange ($2\sigma$) curves in Fig.~\ref{fig:limits}. We join the points of different $g$ values to illustrate the trend of the $M_0/H_*$ bound as a function of $g$. Parameter points that lie below the curves are ruled out. The simple cut-and-count analysis can be sensitive to $M_0$ values that are $\mathcal{O}(100)$ times larger than the Hubble $H_*$ during the inflation. At the same time, the central temperature of the hotspots are determined by a typical value of the heavy mass away from the minimum, and is $\sim 10^4 H_*$. This shows the possibility of looking for extremely heavy particles produced by the inflationary dynamics using our position space search. In the plots we also show, via gray dashed and dotted lines, the bounds that correspond to $C_{\ell}^{\rm TT}$ deviations as given in Table~\ref{tab:CutFlow5}, which give us an idea about how power spectrum searches can constrain the heavy scalar masses. 

We see both for $\eta_*=100$ and 160~Mpc, the position space search performs better. We also show in light blue various contours labelling how many hotspots are produced for different values of $M_0$ and $g$. These curves illustrate the exponential sensitivity in the number of hotspots to $M_0/H_*$, so that slight differences in the bounds -- either comparing the position space search to the power spectrum or jumping $1\sigma \leftrightarrow 2\sigma$ -- correspond to big differences ($\sim$ order of magnitude in some places) in the number of hotspots. For all the bounds, we stay above the ``\% Backreaction'' contour, implying the backreaction on slow-roll dynamics due to particle production is sub-percent. Finally, in the bottom right panel of Fig.~\ref{fig:limits}, we combine different $\eta_*$ benchmarks to see that the bounds do not vary drastically as we change $\eta_*$.

\section{Conclusion}\label{sec.concl}
In this work we have studied how production of very heavy particles during inflation can lead to localized perturbations in the spacetime curvature, and eventually lead to anomalously hot or cold spots on the CMB sky. The key ingredient that leads to heavy particle production is an inflaton-dependent mass of the heavy field $\sigma$ which becomes small (but still $\gg H_*$) at a particular (conformal) time $\eta_*$, i.e., $M^2(\eta) = M_0^2 + f(\phi)$ where $\phi$ is the inflaton, and $M^2(\eta_*) = M_0^2 \ll f(\phi)$ for typical values of $\phi$. Quite generally when the heavy mass pass through such a minimum, we can approximate the mass-squared function as quadratic in $\phi$, $M^2(\eta)\approx g^2(\phi-\phi_*)^2+M_0^2$. This implies the $\sigma-\phi$ coupling is controlled by a single parameter $g$. At $\eta_*$, deviations from adiabaticity are $\mathcal O(H_*/M_0)$, and following the steps laid out in Ref.~\cite{Kofman:1997yn}, we find particle production can be significant even for $M_0 \sim O(100H_*)$. We also compute the number of hotspots that appear within the CMB surface. The non-adiabatic nature of particle production from a time-dependent vacuum also implies that such spots come in pairs with the mutual separation dependent on the size of the comoving horizon $\eta_*$. 

We then discuss in detail how the primordially produced heavy particles lead to temperature perturbations in the CMB, being careful to include subhorizon evolution dominated by Sachs-Wolfe and integrated Sachs-Wolfe effects. Depending on $\eta_*$, these effects can be significant, even turning an initial cold spot into a hotspot or vice-versa, and in particular, our benchmarks choices of $\eta_*$ lead to hotspots. Compared with a previous work involving such primordial particle production~\cite{Flauger:2016idt}, the particle production in our model occurs only at a particular time $\eta_*$, rather than periodically. We also show that the pairwise nature of particle production leads to an overlap of the two hotspots originating from the same pair, and this leads to a non-circular combined hotspot profile.

Varying $M_0/H_*$, $g$ and the production time $\eta_*$, we explore how the hotspot signal can be searched for, both using the CMB power spectrum and with a position-space search that selects out particularly hot pixels. We find the sensitivity for both methods is similar, but the position-space search is more powerful when $g$ is large~($\sim{O}(3-6)$) and for larger $\eta_*\gtrsim 100~$Mpc -- scenarios where the spots are larger and hotter. This sensitivity is better than the one suggested by Ref.~\cite{Munchmeyer:2019wlh} which used an optimized $N$-point analysis rather than a position-space search, though direct comparison is somewhat tricky since they work in the context of a model with periodic particle production. Furthermore, the optimal estimator in Ref.~\cite{Munchmeyer:2019wlh} assumes no overlap between different hotspot profiles, a feature that is absent in our model and analysis. It would be very interesting to see how an optimized $N$-point analysis performs for non-periodic particle production and with overlapping hotspot profiles.

There remains several important and interesting future directions. We have not considered in detail the effects on non-Gaussianity coming from such events of particle production. The resulting bispectrum or trispectrum would exhibit ``bump'' features on specific scales, and therefore new templates of shape function would be needed to detect/constrain such kinds of non-Gaussianity.

The position space search we employ is a first step and involves several approximations. Most significantly, we assume the relevant astrophysical point sources have been masked and instrumental noise has been subtracted out from the CMB map. We believe that aspects of our signal, such as the perfectly circular nature of individual hotspot profiles in the CMB, and their appearance in a limited signal frequency range would help differentiate them from normal astrophysical point sources, though this deserves further study. Our position-space study is also simplistic and does not take advantage of the profile shape (which we have computed theoretically). Further analysis, better exploiting the signal properties, as well as an analysis using wavelet analysis and machine learning to better pick out the signal over the CMB background, is underway and will be discussed in a future work.

\acknowledgments
Some of the results in this paper have been derived using the \texttt{HEALPix} package~\cite{Gorski:2004by}. We thank Daniel Carney, Xingang Chen, Majid Ekhterachian, Raphael Flauger, Daniel Green, Moritz Münchmeyer, Vivian Poulin, Chang Sub Shin and Raman Sundrum for useful conversations. We also thank Moritz Münchmeyer for useful comments on the draft. JK was supported by the National Research Foundation of Korea (NRF) grant funded by the Korea government (MSIT) (No. 2021R1C1C1005076). SK was supported in part by the National Science Foundation (NSF) grant PHY-1915314 and the U.S. Department of Energy Contract DE-AC02-05CH11231. AM was supported in part by the NSF Grant PHY-1820860. YT was supported by the NSF grant PHY-2014165. SK and YT were also supported in part by the NSF grant PHY-1914731 and by the Maryland Center for Fundamental Physics.

\begin{appendix}
\section{Computation of primordial curvature perturbation due to a massive particle}\label{app:profile}
Here we compute the curvature perturbation using the ``in-in'' formalism~\cite{Weinberg:2005vy}.
We start with Eq.~\eqref{eq:tadpole} for the action of a massive particle,
\begin{equation}
S_{\rm particle}\supset-\int_{\eta_*}^0 d\eta\,\partial_{\eta}\zeta\,\frac{M(\eta)}{H_*}\,.
\end{equation}
Here, $\eta_*$ is the (conformal) time of particle production. Assuming the particle is located at $\vec{x}_{\rm HS}$, the above can be rewritten as,
\begin{align}
S_{\rm int}  = - \int_{\eta_*}^0 d\eta \int \frac{d^3\vec{k}}{(2\pi)^3}\frac{M(\eta)}{H_*}\partial_\eta \zeta_{\vec{k}} e^{i\vec{k}\cdot \vec{x}_{\rm HS}}.
\end{align}
The one point function can then be computed using the ``in-in'' formalism where the master formula for evaluating a correlator $Q$ is given by,
\begin{align}
    \langle\Omega|U(t_f,t_i)^\dagger Q U(t_f,t_i)|\Omega\rangle = \langle 0|\Bar{T}e^{+i\int_{-\infty(1+i\epsilon)}^{t_f}dt_2 H_I^{\rm int}(t_2)}Q_I(t_f) Te^{-i\int_{-\infty(1-i\epsilon)}^{t_f}dt_1 H_I^{\rm int}(t_1)}|0\rangle.
\end{align}
Here $T(\Bar{T})$ denote time (anti-time) ordering operators and $H_I^{\rm int}$ is the interacting part of the Hamilotian evaluated in the interaction picture. In the present case, $H_I^{\rm int}$ is obtained from $S_{\rm int}$. Finally, $U(t_f,t_i)$ is the time evolution operator starting from the (interacting) Bunch-Davies vacuum $|\Omega\rangle$ at time $t_i$ and evolving till the final time $t_f$. The standard $i\epsilon$-prescription projects $|\Omega\rangle$ onto the free vacuum $|0\rangle$.

At the leading order in $M$ using the above master formula we get, 
\begin{align}
\langle \zeta_{\vec{p}}(\eta=0)\rangle = (-i)\int_{\eta_*}^0 d\eta  \int \frac{d^3\vec{k}}{(2\pi)^3}\frac{M(\eta)}{H_*}  \langle \zeta_{\vec{p}} \partial_\eta \zeta_{\vec{k}}\rangle e^{i\vec{k}\cdot \vec{x}_{\rm HS}} + \text{anti-time ordered piece}.
\end{align}
Here, the anti-time ordered piece is just the complex conjugate of the first term. To evaluate the integral, we can use expression of the fluctuation mode $\zeta_{\vec{k}}$ in terms of creation and annihilation operators (see e.g.~\cite{Baumann:2009ds}),
\begin{align}
\zeta_{\vec{k}}=\frac{H_*^2}{\dot{\phi}_0}\frac{1}{\sqrt{2k^3}}\left[(1-ik\eta)e^{ik\eta}a_{\vec{k}}^\dagger+(1+ik\eta)e^{-ik\eta}a_{-\vec{k}}\right]. 
\end{align}
Here $\dot{\phi}_0\approx60^2 H_*^2$ as determined by the primordial scalar power spectrum. Upon simplifying further, we get the final momentum-space expression for the one-point function (relabelling $\vec{p}$ as $\vec{k}$),
\begin{align}
\langle\zeta_{\vec{k}}\rangle = \frac{H_*^4}{\dot{\phi}_0^2}\int_{\eta_*}^0 d\eta \frac{M(\eta)}{H_*} \frac{\eta}{k}\sin(k\eta)e^{-i\vec{k}\cdot\vec{x}_{\rm HS}}.    
\end{align}
The time integral can be done after using approximating the mass around $\eta_*$ as, $M/H_*\approx \frac{g\dot{\phi}_0}{H_*^2}\ln(|\eta_*/\eta|)$ from Eq.~\eqref{timedepmass} and dropping the subleading $M_0$ contribution,
\begin{align}\label{eq.zetak}
\langle\zeta_{\vec{k}}\rangle = \frac{gH^2}{\dot{\phi}_0}\frac{1}{k^3}\left(\text{Si}(|k\eta_*|)-\sin(|k\eta_*|)\right)e^{-i\vec{k}\cdot\vec{x}_{\rm HS}},     
\end{align}
where $\text{Si}(x)=\int_0^x dt \sin(t)/t$. The above approximation for $M$ is valid as long as $\eta$ is within a few e-foldings of $\eta_*$, and will be sufficient for our purpose. It is also useful to get the position-space answer~\cite{Maldacena:2015bha}, 
\begin{align}
\langle\zeta(\vec{x})\rangle=&\frac{H_*^4}{\dot{\phi}_0^2}\int \frac{d^3\vec{k}}{(2\pi)^3}e^{i\vec{k}\cdot (\vec{x}-\vec{x}_{\rm HS})}  \int_{\eta_*}^0 d\eta \frac{M(\eta)}{H_*} \frac{\eta}{k}\sin(k\eta),\nonumber\\  
=&\frac{H_*^4}{\dot{\phi}_0^2}\frac{1}{4\pi^2}\int_{0}^{\infty} dk k \frac{e^{ikr}-e^{-ikr}}{ikr}\int_{\eta_*}^0 d\eta \frac{M(\eta)}{H_*}\eta \sin(k\eta),\nonumber\\
=&-\frac{H_*^4}{\dot{\phi}_0^2}\frac{1}{4\pi r}\int_{\eta_*}^0 d\eta \eta \frac{M(\eta)}{H_*}\delta(\eta+r),\nonumber\\
=&\frac{H_*^4}{\dot{\phi}_0^2}\frac{M(\eta=-r)}{4\pi H_*},
\end{align}
where $r=|\vec{x}-\vec{x}_{\rm HS}|$. Finally, using the expression $\dot{\phi}_0^2=2\epsilon H_*^2 \mpl^2$ and noting the fact the particle production happens only around $\eta_*$, we get Eq.~\eqref{eq:hsprofile}.

\section{Additional images of the PHS}\label{app.img}
\begin{figure}[h]
    \centering
    \includegraphics[width=0.45\textwidth,clip]{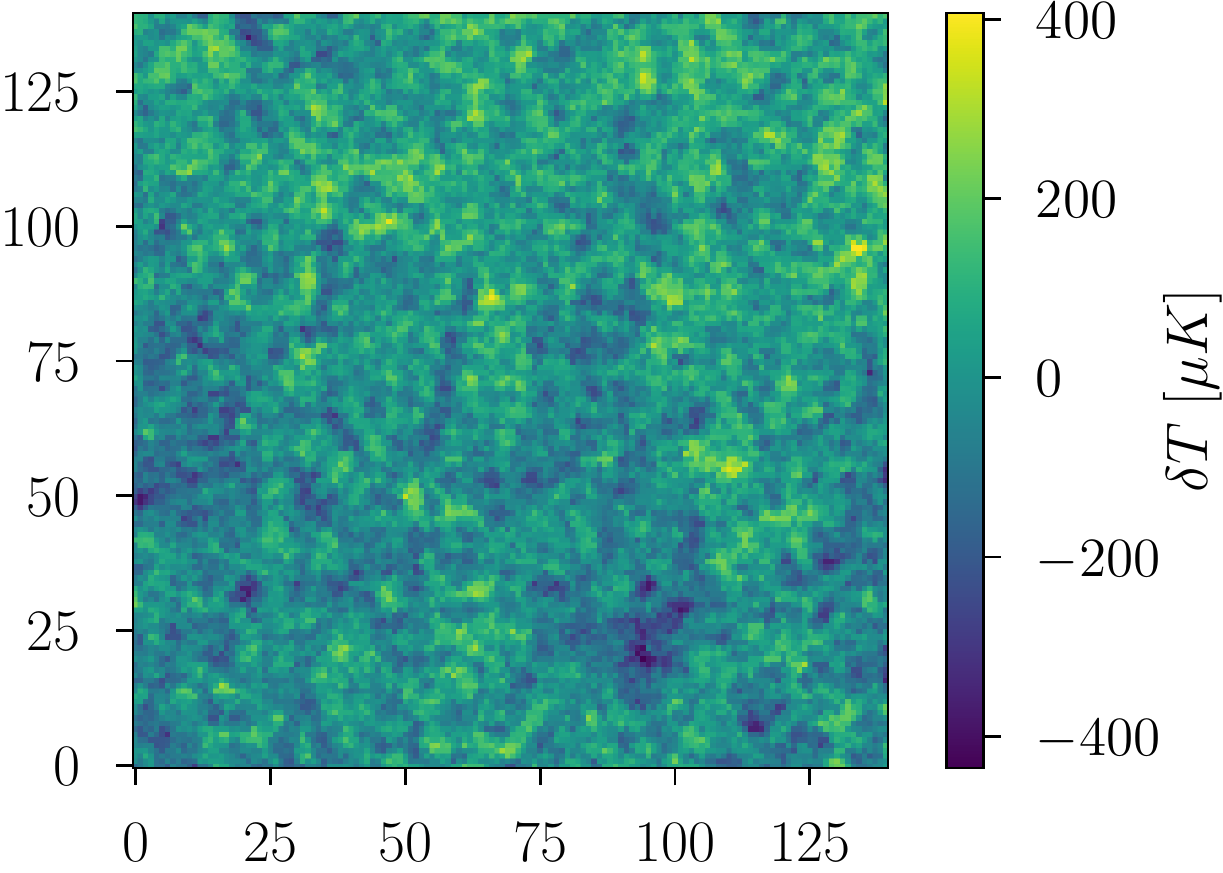}\quad
    \includegraphics[width=0.437\textwidth,clip]{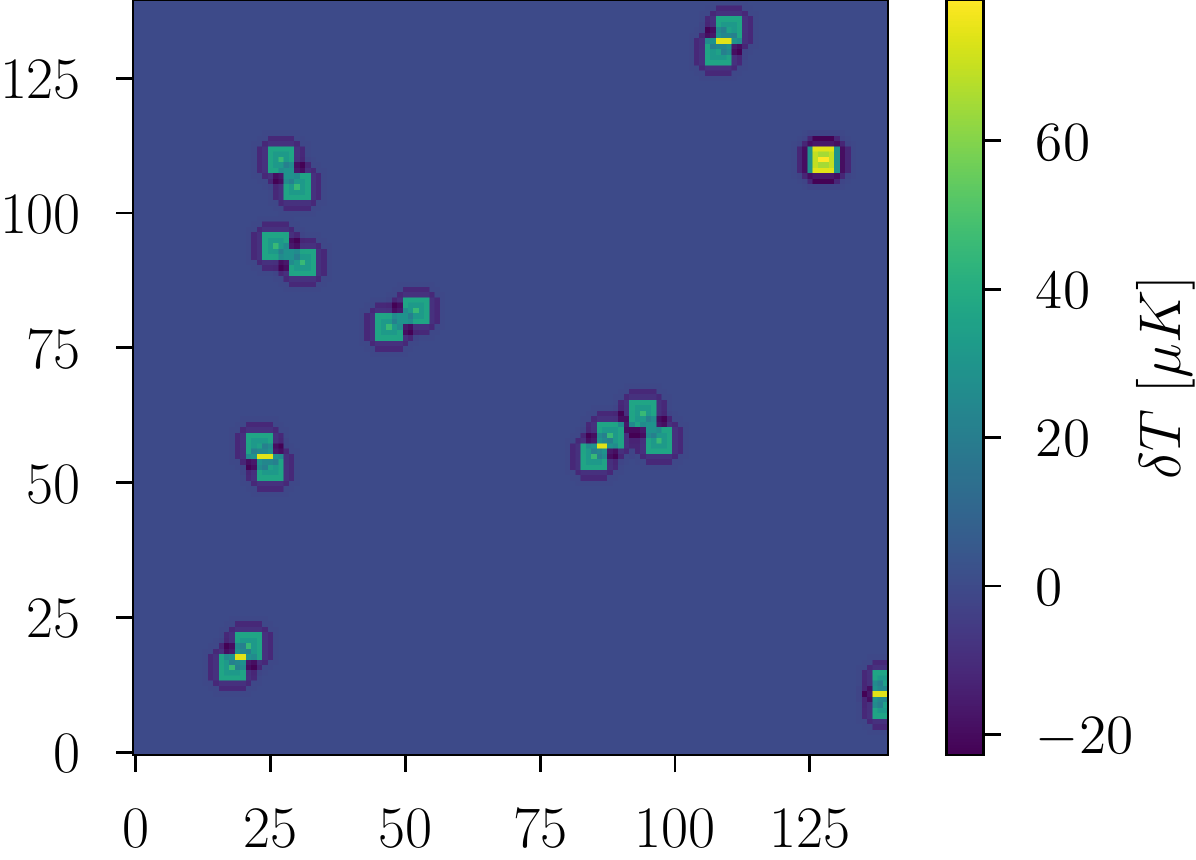}\\
    \includegraphics[width=0.45\textwidth,clip]{Plots/Low_Resol_BKG_Nside256.pdf}   \quad
    \includegraphics[width=0.437\textwidth,clip]{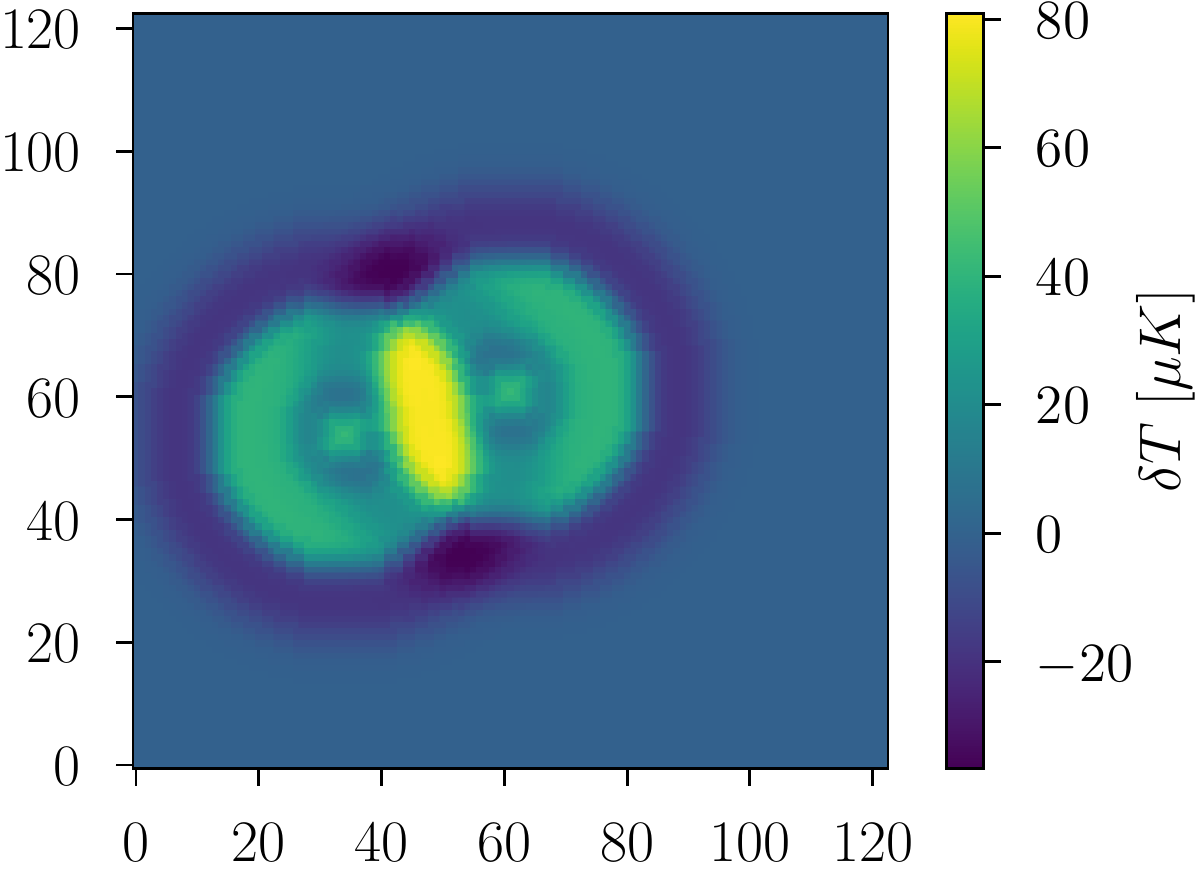}  
    \caption{Local patches of (upper-left) CMB + PHS, (upper-right) PHS, and (lower-left) CMB, generated within a region of longitude = $ [-\ang{16}, \ang{16}$] and latitude = $[-\ang{16}, \ang{16}]$ centered at $(\theta, \phi) = (\pi/2,0)$ and projected onto the Cartesian coordinates ($140 \times 140$ pixels) using $N_{\rm side} = 256$.  The shape of PHS is based on the $\eta_{*} = 100$ Mpc mode using $g =6$, and the separation distance is $r_{\text{sep}} = 0\sim6$ pixels. In total, ten PHS signals are injected using a flat distribution. (lower-right) A higher-resolution image of the PHS using $N_{\rm side} = 2048$. 
    }
    \label{fig:PHSsignal2}
\end{figure}

\begin{figure}
    \centering
    \includegraphics[width=0.45\textwidth,clip]{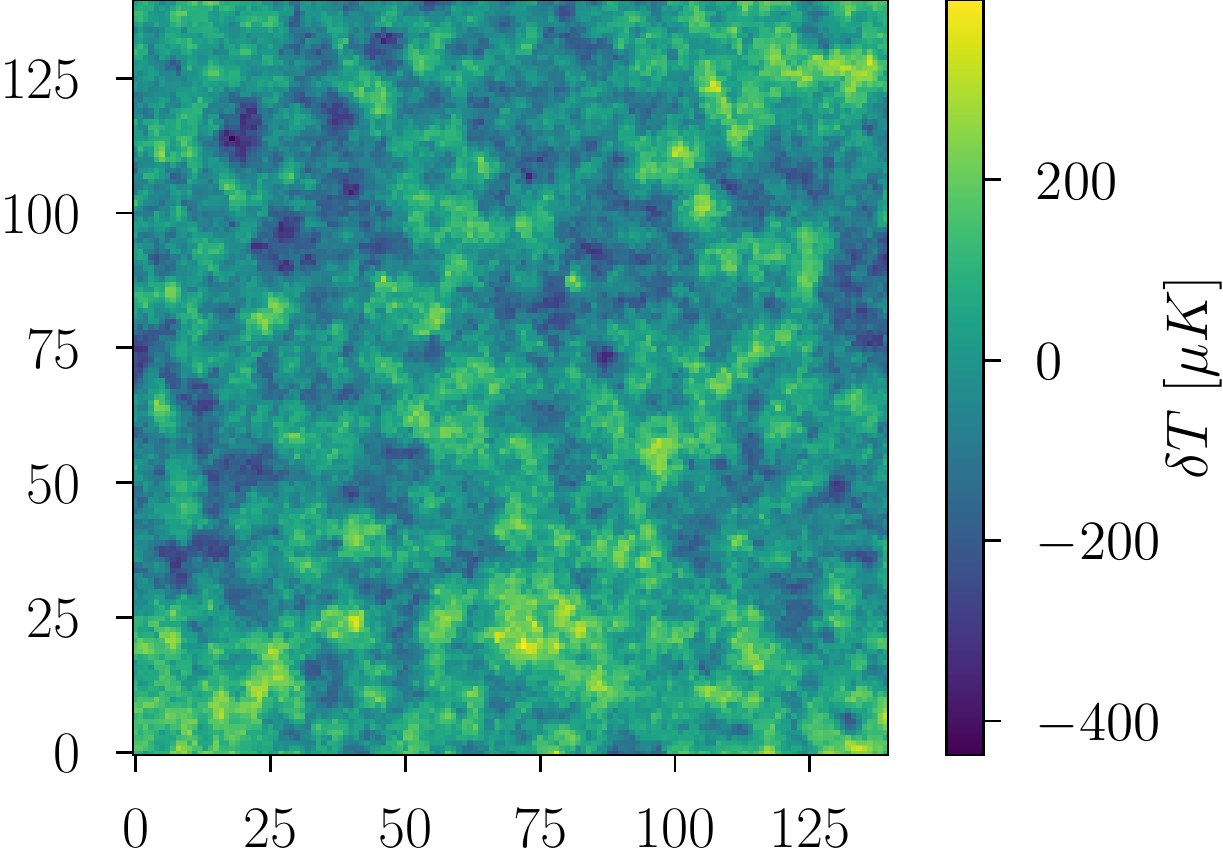}\quad
    \includegraphics[width=0.437\textwidth,clip]{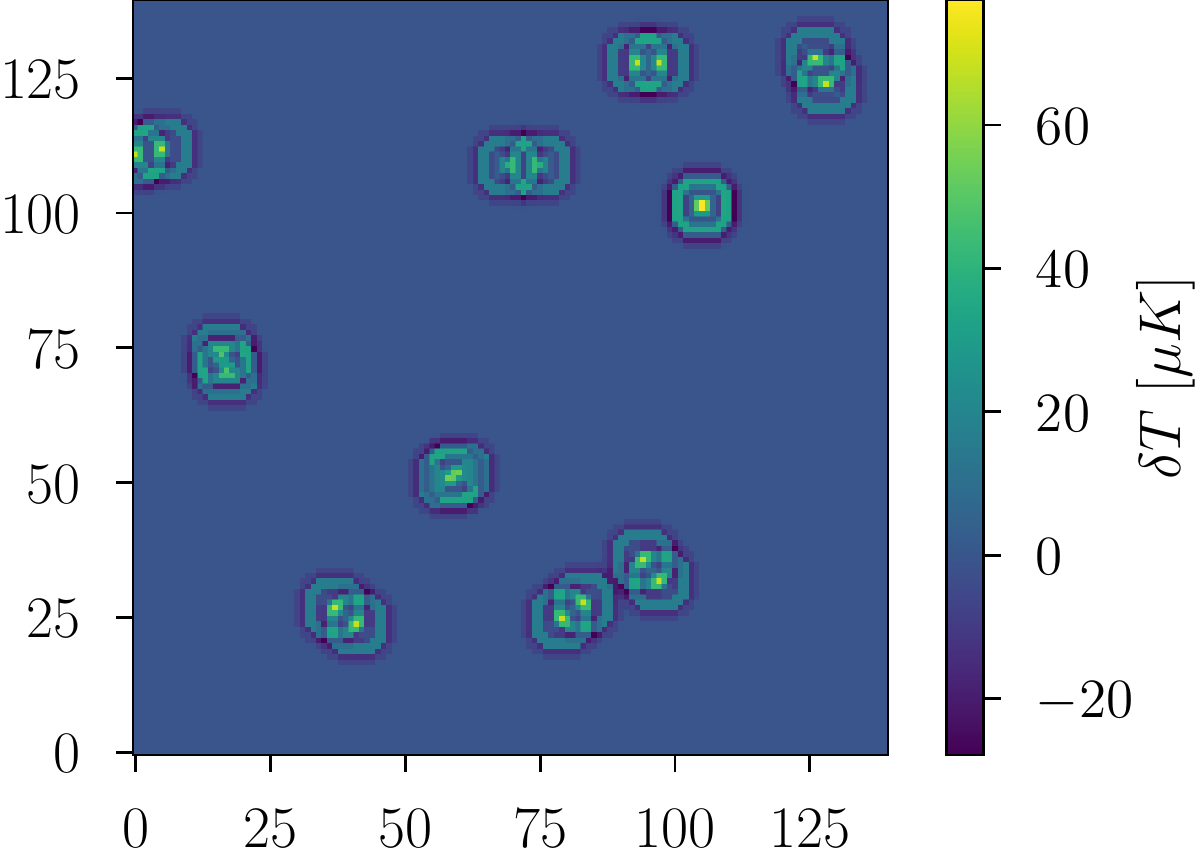}\\
    \includegraphics[width=0.45\textwidth,clip]{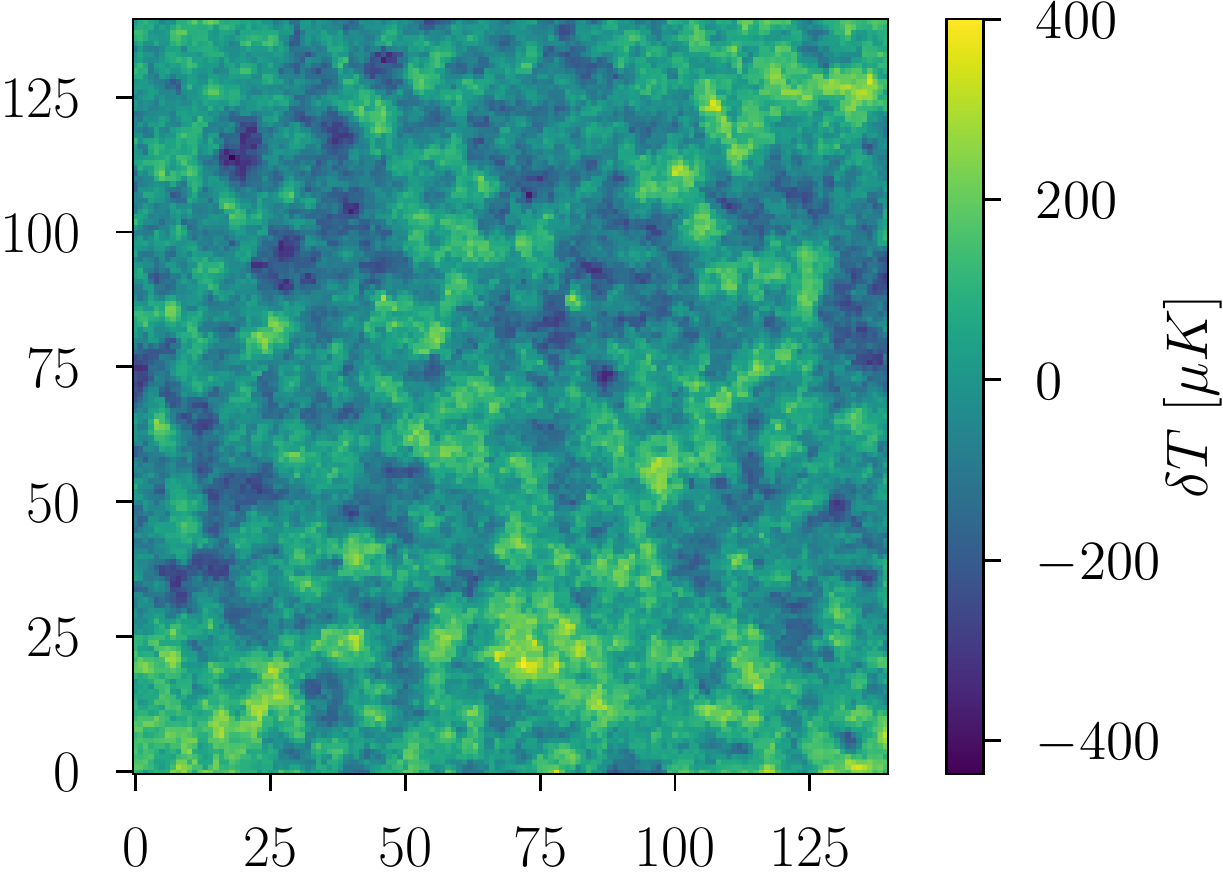}   \quad
    \includegraphics[width=0.437\textwidth,clip]{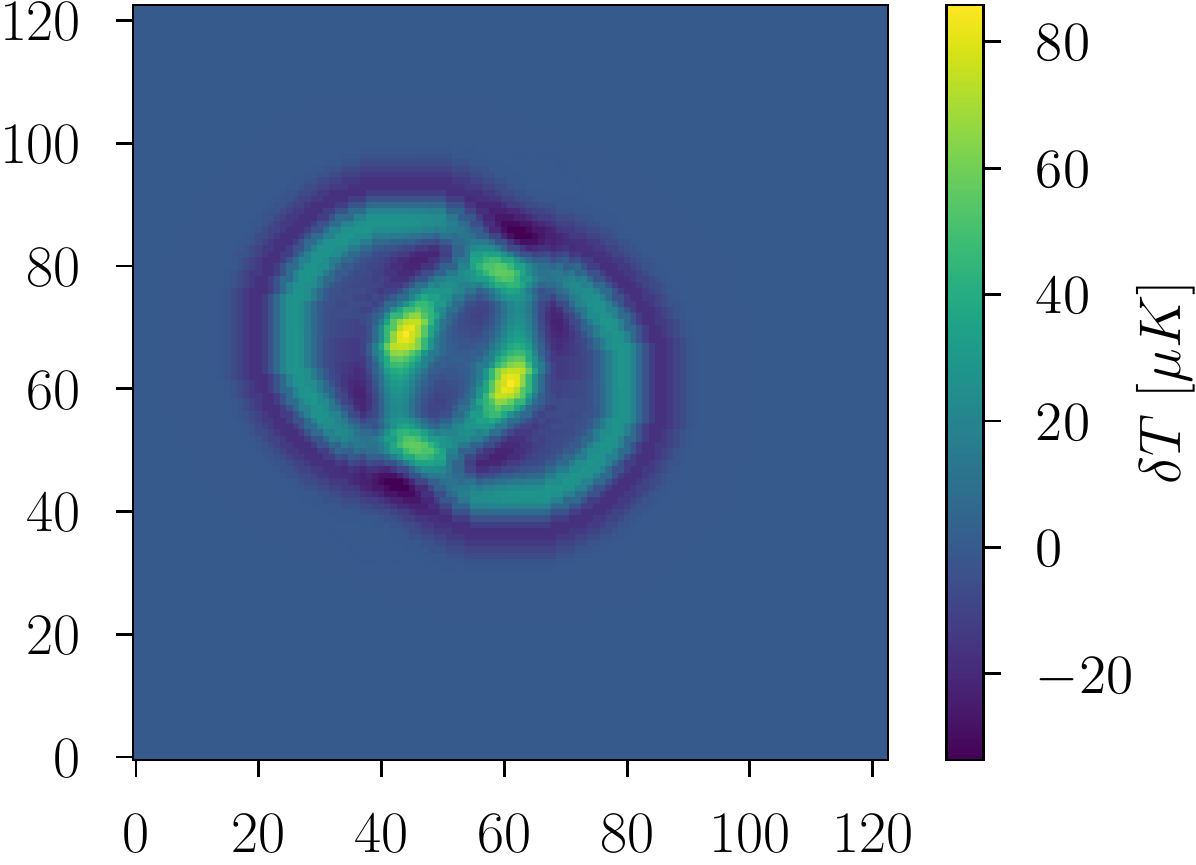}  
    \caption{Local patches of (upper-left) CMB + PHS, (upper-right) PHS, and (lower-left) CMB, generated within a region of longitude = $ [-\ang{8}, \ang{8}$] and latitude = $[-\ang{8}, \ang{8}]$ centered at $(\theta, \phi) = (\pi/2,0)$ and projected onto the Cartesian coordinates ($140 \times 140$ pixels) using $N_{\rm side} = 512$.  The shape of PHS is based on the $\eta_{*} = 50$ Mpc mode using $g =10$, and the separation distance is $r_{\text{sep}} = 0\sim6$ pixels. In total, ten PHS signals are injected using a flat distribution. (lower-right) A higher-resolution image of the PHS using $N_{\rm side} = 2048$.  }
    \label{fig:PHSsignal3}
\end{figure}

The images of PHS for the $\eta_*=100~\text{Mpc},\,50~\text{Mpc}$ benchmark scenarios are shown respectively in Figs.~\ref{fig:PHSsignal2} and~\ref{fig:PHSsignal3}.


\end{appendix}
\bibliographystyle{JHEP}
\bibliography{references}
\end{document}